\documentclass[aps,prxquantum,reprint,groupedaddress,amsmath,amssymb,longbibliography]{revtex4-2}

\usepackage{CJK}
\usepackage{graphicx} 
\usepackage{bm}
\usepackage[pdfstartview=FitH, CJKbookmarks=true, bookmarksnumbered=true, bookmarksopen=true, colorlinks=true, pdfborder=0 0 1, citecolor=blue, linkcolor=blue, linktocpage=true] {hyperref}
\usepackage{epstopdf} 

\begin{document}

\title{Calculations of Chern number: equivalence of real-space and twisted-boundary-condition formulae}

\author{Ling Lin$^{1,2}$}
\author{Yongguan Ke$^{3}$}
\author{Li Zhang$^{1,2}$}
\author{Chaohong Lee$^{1,2,4}$}
\email{Email: chleecn@szu.edu.cn, chleecn@gmail.com}

\affiliation{$^{1}$Institute of Quantum Precision Measurement, State Key Laboratory of Radio Frequency Heterogeneous Integration, Shenzhen University, Shenzhen 518060, China}
\affiliation{$^{2}$College of Physics and Optoelectronic Engineering, Shenzhen University, Shenzhen 518060, China}
\affiliation{$^{3}$Laboratory of Quantum Engineering and Quantum Metrology, School of Physics and Astronomy, Sun Yat-Sen University (Zhuhai Campus), Zhuhai 519082, China}
\affiliation{$^{4}$Quantum Science Center of Guangdong-Hongkong-Macao Greater Bay Area (Guangdong), Shenzhen 518045, China}

\date{\today}

\begin{abstract}
Chern number is a crucial invariant for characterizing topological feature of two-dimensional quantum systems.
Real-space Chern number allows us to extract topological properties of systems without involving translational symmetry, and hence plays an important role in investigating topological systems with disorder or impurity.
On the other hand, the twisted boundary condition (TBC) can also be used to define the Chern number in the absence of translational symmetry.
Based on the perturbative nature of the TBC under appropriate gauges, we derive the two real-space formulae of Chern number (namely the non-commutative Chern number and the Bott index formula), which are numerically confirmed for the Chern insulator and the quantum spin Hall insulator.
Our results not only establish the equivalence between the real-space and TBC formula of the Chern number, but also provide concrete and instructive examples for deriving the real-space topological invariant through the twisted boundary condition.
\end{abstract}


 
\maketitle
 
\section{Introduction}
Topological band theory has achieved great success in characterizing the topological nature of quantum matters in the past decades \cite{RevModPhys.82.3045, RevModPhys.83.1057, kane2013topological, RevModPhys.88.035005}.
Beyond the well-known spontaneous breaking theory, topological quantum states are classified by intrinsic topological invariants \cite{PhysRevLett.49.405,RevModPhys.88.035005}.
In the context of band theory, topological invariants are usually defined on the Bloch manifold.
A fascinating nature of topological quantum state is that it is immune to perturbations, i.e. disorder and impurity, provided that the spectral gap and the underlying symmetry are preserved.
However, the translational symmetry will be broken in a spatially disordered system, invalidating the band structure.
In this situation, the topological band theory fails to give the topological invariant in systems without translational symmetry.

To characterize the topological nature in the presence of disorder, one may use the twisted boundary condition (TBC) to formulate topological invariants~\cite{Niu_1984, PhysRevB.30.1097, PhysRevB.31.3372,RevModPhys.82.1959,PhysRevX.8.031079,PhysRevB.103.224208}.
One of the prominent features of the TBC is that it operates in real space.
This method works in quasi-periodic boundary condition and does not require any translational symmetry.
Physically, topological invariants are closely related to the system's response of current to the external perturbation.
The TBC is equivalent to piercing magnetic flux through the system \cite{PhysRevB.23.5632,PhysRevB.30.1097,PhysRevLett.96.060601}, which results in adiabatic currents in the system \cite{PhysRevLett.68.1375}.
Hence, one is able to compute the corresponding response of current in the system to obtain the topological invariant, e.g. the quantum Hall conductance \cite{PhysRevB.31.3372,PhysRevB.30.1097,PhysRevLett.54.259} and polarization \cite{Niu_1984, niu1991theory, RevModPhys.82.1959, PhysRevX.8.021065, PhysRevLett.128.246602}.
The twist angle is related to the quasi momentum of the eigenstate under certain gauge choice in disorder-free system, connecting the topological invariant defined through quasi momenta and twist angles \cite{PhysRevB.107.125161}.
Recently, the real-space formula of topological invariant draws many attentions.
Unlike the momentum-space topological invariant that are defined on Brillouin manifold, one can construct topological invariants directly in real space, allowing computation without translational symmetry.
Roughly speaking, the real-space formula of topological invariant can be classified into two types, (i) the non-commutative form \cite{PhysRevLett.105.115501,PhysRevB.80.125327, Prodan_2010, Prodan_2011, prodan2017computational} and (ii) the Bott index form \cite{exel1991invariants, Loring_2010, hastings2010almost,HASTINGS20111699, loring2015k}.
In 1D topological insulators with chiral symmetry, the winding number can be well expressed by these two methods in real space~\cite{PhysRevLett.113.046802, PhysRevB.103.224208}.
Moreover, in 1D systems, they are proved to be equivalent to the winding number defined through the TBC in the thermodynamic limit, and hence these two real-space formulae are also equivalent \cite{PhysRevB.103.224208}.
In 2D Chern insulator, the real-space Chern number is also found to possess the non-commutative form and the Bott index form.
Both methods have been extensively used in studying disorder effect to topological insulators.
However, it is still unclear whether the real-space formulae of Chern number can be related to the Chern number defined via the TBC.
In this article, we show that the Chern number defined via the TBC and the two real-space formulae (the non-commutative method and the Bott index method) are equivalent in the thermodynamic limit.
The key point of the proof is to make use of the perturbative nature of the twist angles in TBC and expand operators and eigenstates up to the first order of $\theta_j/L_j$ with $L_j$ being the number of cells in $j=x,y$ direction.
With the perturbative analysis, the Chern number defined via TBC is reduced to the real-space formulae of Chern number.
This provides an useful approach to attain the real-space formula of topological invariants.
We also show that the real-space formula of the spin Chern number in quantum spin Hall (QSH) insulator can be also derived from a generalized TBC.
By imposing different generalized TBCs, one can attain different real-space formulae.
In addition, we note that the Berry curvature defined via TBC becomes constant in the thermodynamic limit and away from the phase boundary.
In finite systems, the flatness of the Berry curvature will affect the accuracy of the non-commutative method, while the Bott index method is immune to the non-flat effects.
To verify our arguments, we present numerical calculations for the Haldane model and the Kane-Mele model.
We compute the flatness of Berry curvature defined via TBC, and show that it tends to be a constant in the thermodynamic limit except for the phase transition point.
We also numerically verify the equivalence of Chern number given by non-commutative method and the Bott index method.
The rest of this paper is organized as follows:
In Sec.~\ref{sec:Chern_number_TBC}, we briefly introduce the Chern number defined via the TBC. 
We consider a non-interacting system and target gapped eigenstates.
%
%
In Sec.~\ref{sec:Numerical_Verification}, we perform numerical calculations for the Haldane model and Kane-Mele model to verify our arguments, including the flatness of Berry curvature and the calculation of Chern number in real space.
In Sec.~\ref{sec:Discussion}, we briefly summarize and discuss the results.

\begin{figure}[htp]
\centering
  \includegraphics[width = 0.9\columnwidth ]{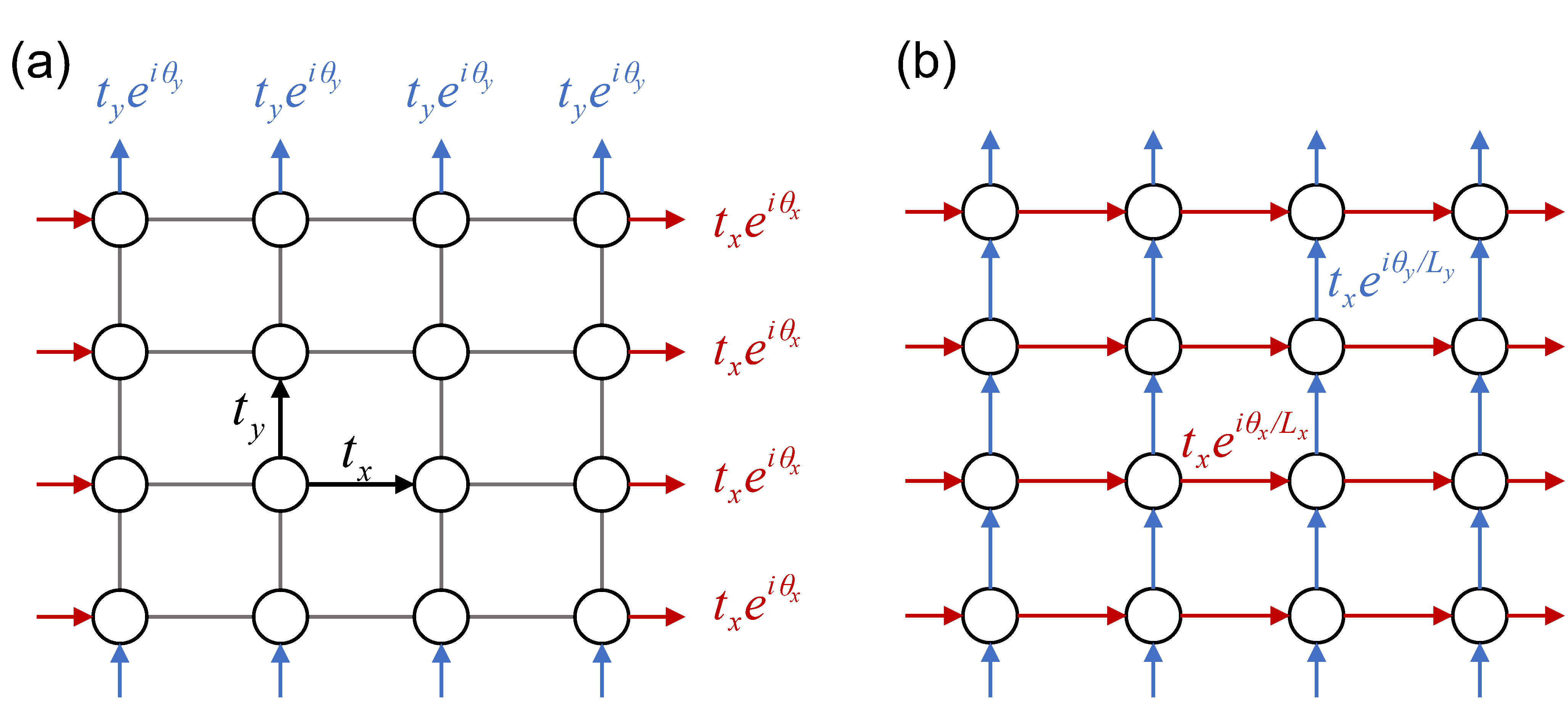}
  \caption{\label{fig:FIG_TBC_illustration}
  Schematic illustration of the TBC in a 2D simple lattice.
  In both configurations, edges are glued together in the same manner as the periodic boundary condition.
  (a) is a general form of the TBC. The gauge field only appears at the boundary, and the particle gains extra phase when it tunnels through the boundary, as indicated by color arrows.
  (b) is transformed from (a) via a large gauge transformation. 
  The gauge field distributes uniformly throughout the bulk, and the field strength is diluted.
    }
\end{figure}

\section{Chern number: from twisted boundary condition to real-space formulae}
\label{sec:Chern_number_TBC}

\subsection{Chern number defined via twisted boundary condition}

In this section, we introduce the TBC and the Chern number defined through TBC.
Under the TBC, the lattice's edges are ``glued" together pairwise as a 2D torus, just like the periodic boundary condition (PBC).
The core is that we enforce particles to gain an extra phase $\boldsymbol{\theta}$ after they tunnel through the boundary, as demonstrated in Fig.~\ref{fig:FIG_TBC_illustration} (a).
Apparently, $\boldsymbol{\theta} = 0$ corresponds exactly to the PBC.
Next, we consider a non-interacting 2D system under TBC.
The system's Hamiltonian satisfies $\hat H\left( {{\theta _x} + 2\pi ,{\theta _y}} \right) = \hat H\left( {{\theta _x},{\theta _y} + 2\pi } \right) = \hat H\left( {{\theta _x},{\theta _y}} \right)$.
Then, we target a set of single-particle eigenstates that are gapped to other eigenstates.
Note that, if translation symmetry is preserved, these eigenstates correspond to gapped bands.
For simplicity, we collect these eigenstates and introduce the following notation
\begin{equation}
\label{eqn:spinor_representation}
{{\boldsymbol{\Psi }}_{\boldsymbol{\theta }}} = \left( {\begin{array}{*{20}{c}}
{|{\psi _1}\left( {\boldsymbol{\theta }} \right)\rangle }&{|{\psi _2}\left( {\boldsymbol{\theta }} \right)\rangle }& \cdots &{|{\psi _{\cal N}}\left( {\boldsymbol{\theta }} \right)\rangle }
\end{array}} \right),
\end{equation}
where $|\psi_\mu \left( {\boldsymbol{\theta }} \right)\rangle$ is the $\mu$-th eigenstate of the Hamiltonian under TBC, and $\mathcal{N}$ is the total number of these targeted states.
There is ${{\boldsymbol{\Psi }}^\dag_{\boldsymbol{\theta }}} {{\boldsymbol{\Psi }}_{\boldsymbol{\theta }}}  = I_{\mathcal{N}}$ with $ I_{\mathcal{N}}$ being the $\mathcal{N} \times \mathcal{N}$ identity matrix,  and the projector of this subspace can be expressed as ${{\boldsymbol{\Psi }}_{\boldsymbol{\theta }}} {{\boldsymbol{\Psi }}^\dag_{\boldsymbol{\theta }}}  = \sum_{\mu}{|{\psi _\mu }\left( {\boldsymbol{\theta }} \right)\rangle \langle {\psi _\mu }\left( {\boldsymbol{\theta }} \right)|}$.
These targeted states are assumed to remain gapped for $\theta_x, \theta_y \in [0, 2\pi]$, which is true in the thermodynamic limit.
The Chern number of these targeted states can be expressed through the twist angles $\left( {{\theta _x},{\theta _y}} \right)$ \cite{PhysRevB.31.3372}:
\begin{equation}
\label{eqn:ChernNum_TBC}
{C_{{\mathrm{TBC}}}} = \frac{1}{{2\pi }}\int\limits^{2\pi }_{0} {\int\limits^{2\pi }_{0} {{\mathrm{Tr}}\left[ {\mathcal{F}\left( {\boldsymbol{\theta }} \right)} \right]} } {\mathrm{d}^2 \boldsymbol{\theta} },
\end{equation}
in which ${\mathcal F}\left( {\boldsymbol{\theta}} \right)$ is the non-Abelian Berry curvature
\begin{eqnarray}
\label{eqn:nonAbelian_BerryCurvature_TBC}
{\mathcal F}\left( {\boldsymbol{\theta}} \right) &= &  {\partial _{{\theta}_x}}{{\mathcal A}_y}\left( {\boldsymbol{\theta}} \right) - {\partial _{{\theta}_y}}{{\mathcal A}_x}\left( {\boldsymbol{\theta}} \right) +\left[ {{{\mathcal A}_x}\left( {\boldsymbol{\theta}} \right),{{\mathcal A}_y}\left( {\boldsymbol{\theta}} \right)} \right], \nonumber \\
{\mathcal{A}_j}\left( {\boldsymbol{\theta }} \right) &=& -i{{\boldsymbol{\Psi }}^\dag_{\boldsymbol{\theta }}}{\partial _{{\theta _j}}}{\boldsymbol{\Psi }}_{\boldsymbol{\theta }} ,\;\quad j = x,y.
\end{eqnarray}
Note that ${\mathcal{A}_j}\left( {\boldsymbol{\theta }} \right) $ is a $\mathcal{N}$-by-$\mathcal{N}$ matrix.
The minus sign in ${\mathcal{A}_j}\left( {\boldsymbol{\theta }} \right) $ is related to the sign of twist angles defined in the TBC.
Eq.~\eqref{eqn:ChernNum_TBC} can well capture the topological property (quantum Hall conductance) of the system \cite{PhysRevB.31.3372}.
%
%

Alternatively, by introducing the projector that projects onto the targeted states ${\hat{P}_{\boldsymbol{\theta }}} = {{\boldsymbol{\Psi }}_{\boldsymbol{\theta }}} {{\boldsymbol{\Psi }}^\dag_{\boldsymbol{\theta }}}$, the Chern number can be equivalently expressed as \cite{PhysRevLett.51.51}
\begin{equation}
\label{eqn:ChernNum_TBC_Proj_expression}
C_{\mathrm{TBC}} = \frac{1}{{2\pi i}}\int\limits_0^{2\pi } {\int\limits_0^{2\pi } {{\mathrm{Tr}}\left( {{\hat{P}_{\boldsymbol{\theta }}}\left[ {{\partial _{{\theta _x}}}{\hat{P}_{\boldsymbol{\theta }}},{\partial _{{\theta _y}}}{\hat{P}_{\boldsymbol{\theta }}}} \right]} \right)} \mathrm{d}^2{\boldsymbol{\theta}}}  .
\end{equation}
Given that the targeted states are gapped, the projector will finally return to itself after the twist angle changes $2\pi$, that is: $\hat{P}_{\boldsymbol{\theta  } + 2\pi \hat{\bold{e}}_j } = {\hat{P}_{\boldsymbol{\theta }}}$, with $\bold{e}_j$ the unit vector along $j=x,y$ direction.
In other words, the projector is a periodic function of twist angles.

\subsection{Large gauge transformation and uniform gauge}
\label{sec:Perturbative_expansion}
Under the TBC, the extra phase gained by particle at the boundary can be regarded as a result of gauge field, and therefore there exists gauge freedom.
It is beneficial to transform the gauge field at the boundary to a uniformly distributed gauge field throughout the whole system.
By introducing the twist operator \cite{PhysRevLett.79.1110, PhysRevB.98.155137}
\begin{equation}
{{\hat U}_{\boldsymbol{\theta }}} = \exp \left( {i\frac{{\theta_x }}{{{L_x}}}{{\hat r}_x} + i\frac{{\theta_y }}{{{L_y}}}{{\hat r}_y}} \right),
\end{equation}
where $\hat r_j = \sum_{\boldsymbol{r}} {r_j \hat{n}_{\boldsymbol{r}}}$ is the position operator in $j=x,y$ direction, we can bring the original system under TBC to a unitarily equivalent system via the large gauge transformation
\begin{equation}
\label{eqn:Ham_PBC_gauge_transform}
{\tilde{ H}}\left( {\boldsymbol{\theta }} \right) = {{\hat U}_{\boldsymbol{\theta }}}{{\hat H}}\left( {\boldsymbol{\theta }} \right)\hat U_{\boldsymbol{\theta }}^{ - 1}.
\end{equation}
Here, ${{\tilde{H}}}\left( {\boldsymbol{\theta }} \right) $ indicates the transformed Hamiltonian.
The two unitarily equivalent systems are illustrated in Fig.~\ref{fig:FIG_TBC_illustration}.
To distinguish these two gauges, we use the tilde notation to stress that the operator (eigenstate) is under the gauge that the twist angles uniformly distribute.
For convenience, we will call this gauge the \emph{uniform gauge}, while the original TBC is called \emph{boundary gauge}.
In this case, the Chern number can still be written as the same form
\begin{equation}
\label{eqn:ChernNum_TPG}
\tilde{C}_{\mathrm{TBC}} = \frac{1}{{2\pi }}\int\limits^{2\pi }_{0} { \int\limits^{2\pi }_{0} {\mathrm{Tr}}\left[ {\tilde{\mathcal{F}}\left( {\boldsymbol{\theta }} \right)} \right] }  {\mathrm{d}^2 \boldsymbol{\theta} },
\end{equation}
and
\begin{eqnarray}
\label{eqn:nonAbelian_BerryCurvature_TPG}
\tilde{\mathcal F}\left( {\boldsymbol{\theta}} \right) &= &{\partial _{{\theta}_x}}{\tilde{\mathcal A}_y}\left( {\boldsymbol{\theta}} \right) - {\partial _{{\theta}_y}}{\tilde{\mathcal A}_x}\left( {\boldsymbol{\theta}} \right) + \left[ {{\tilde{\mathcal A}_x}\left( {\boldsymbol{\theta}} \right),{\tilde{\mathcal A}_y}\left( {\boldsymbol{\theta}} \right)} \right], \nonumber \\
{\tilde{\mathcal{A}}_j}\left( {\boldsymbol{\theta }} \right) &=& -i{{\tilde{\boldsymbol{\Psi }}}^\dag _{\boldsymbol{\theta }}}{\partial _{{\theta _j}}}{\tilde{\boldsymbol{\Psi }}}_{\boldsymbol{\theta }},\;\quad j = x,y.
\end{eqnarray}
Note that $ {{\tilde{\boldsymbol{\Psi }}}_{\boldsymbol{\theta }}} = \hat{U}_{\boldsymbol{\theta}}  {{{\boldsymbol{\Psi }}}_{\boldsymbol{\theta }}} $.
It can be proved that Eq.~\eqref{eqn:nonAbelian_BerryCurvature_TBC} and Eq.~\eqref{eqn:nonAbelian_BerryCurvature_TPG} are equivalent: $\tilde{C}_{\mathrm{TBC}} = {C}_{\mathrm{TBC}}$, see the proof in \ref{appendix:ChernNumber_boundary_periodic_gauge_equivalence}.
Similar to Eq.~\ref{eqn:ChernNum_TBC_Proj_expression}, we can formulate the Chern number through the projector ${\tilde{P}_{\boldsymbol{\theta }}} = {\tilde{\boldsymbol{\Psi }}_{\boldsymbol{\theta }}} {\tilde{\boldsymbol{\Psi }}^\dag_{\boldsymbol{\theta }}}$
\begin{equation}
\label{eqn:ChernNum_PBC_Proj_expression}
\tilde{C}_{\mathrm{TBC}} = \frac{1}{{2\pi i}}\int\limits_0^{2\pi} {\int\limits_0^{2\pi } }{\mathrm{Tr}}\left( {{\tilde{P}_{\boldsymbol{\theta }}}\left[ {{\partial _{{\theta _x}}}{\tilde{P}_{\boldsymbol{\theta }}},{\partial _{{\theta _y}}}{\tilde{P}_{\boldsymbol{\theta }}}} \right]} \right)  {\mathrm{d}^2 \boldsymbol{\theta} } ,
\end{equation}
in which the projector satisfies 
\begin{equation}
\label{eqn:P_theta_tilde_relation}
{\tilde P_{\boldsymbol{\theta }}} = {\hat U_{\boldsymbol{\theta }}}{\hat{P}_{\boldsymbol{\theta }}}{\hat U_{\boldsymbol{\theta }}}^{ - 1}.
\end{equation}

In the transformed Hamiltonian \eqref{eqn:Ham_PBC_gauge_transform}, the twist angle $\theta$ is transformed to distribute in all tunneling terms of the system.
Particles gain a ``diluted" phase factor during the tunneling process, that is
\begin{equation}
{{\hat U}_{\boldsymbol{\theta }}}\hat c_{{\boldsymbol{r}} + {\boldsymbol{d}}}^\dag {c_{\boldsymbol{r}}}{{\hat U}_{\boldsymbol{\theta }}}^{ - 1} = {e^{i\left( {\frac{{{\theta _x}}}{{{L_x}}}{d_x} + \frac{{{\theta _y}}}{{{L_y}}}{d_y}} \right)}}\hat c_{{\boldsymbol{r}} + {\boldsymbol{d}}}^\dag {c_{\boldsymbol{r}}}.
\end{equation}
When the tunneling range is finite $d_{x,y} \ll L_{x,y}$, the twist angle appears as $\theta_j/L_j,\; j=x,y$, which can be considered as a perturbation.
Hence, it is desirable to expand operators or eigenstates up to the first order of $1/L_j$ for sufficiently large system away from the critical point.
This also explains why the gapped eigenstate will remain gapped for arbitrary $\theta_{x,y}$ in the thermodynamic limit $L_{x,y} \to \infty$, as the energy gap is only weakly perturbed.
In next subsections, we shall make use of the perturbative nature of the twist angle to derive the two real-space formulae of the Chern number.

\subsection{Real-space Chern number via non-commutative method}
In this subsection, we derive the  the non-commutative real-space formula of the Chern number.
This formula is firstly proposed in Ref.~\cite{PhysRevLett.105.115501, Prodan_2011}, which is based on the momentum-space formula of the Chern number \cite{bellissard1995noncommutative}
\begin{equation}
\label{eqn:ChernNum_momentum_projector_form}
C = \frac{1}{{2\pi i}}\int\limits_{B.Z.} {{\mathrm{Tr}}\left\{ {\hat{ \mathbb{P}}\left( {\boldsymbol{k}} \right)\left[ {{\partial _{{k_x}}}\hat {\mathbb{P}}\left( {\boldsymbol{k}} \right),{\partial _{{k_y}}}\hat{\mathbb{P}}\left( {\boldsymbol{k}} \right)} \right]} \right\}{{\mathrm{d}}^2}{\boldsymbol{k}}},
\end{equation}
where $\hat{\mathbb{P}}\left( {\boldsymbol{k}} \right) = \sum\nolimits_{n \in {\mathrm{occ}}.} {|u_{\boldsymbol{k}}^n\rangle \langle u_{\boldsymbol{k}}^n|} $ is the projector onto Bloch states of occupied bands with momentum $\boldsymbol{k}$.
To derive the real-space formula, their idea is to transform the partial derivative and the integral in Brillouin zone to real space \cite{PhysRevLett.105.115501, Prodan_2011}, which leads to the following non-commutative form of Chern number
\begin{equation}
\label{eqn:ChernNum_Noncomm_original}
C = - 2\pi i\sum\limits_\alpha  {\langle \boldsymbol{r}_0,\alpha |\hat P\left[ { \left[ {{{\hat r}_x},\hat P} \right], \left[ {{{\hat r}_y},\hat P} \right]} \right]|\boldsymbol{r}_0,\alpha \rangle } ,
\end{equation}
where $\hat{P}$ is the projector onto eigenstates in occupied bands, and $|\boldsymbol{r}_0,\alpha\rangle$ is the real-space basis denoting the cell at $\boldsymbol{r}_0$ and $\alpha$ is the label of internal orbits within the cell.
This formula has been successfully applied to various systems \citep{PhysRevLett.105.115501, Prodan_2013_noncommutative, Bourne_2018}.
Next, we would like to show that the non-commutative Chern number can be equivalently derived through the TBC Chern number [Eq.~\eqref{eqn:ChernNum_PBC_Proj_expression}].
For this purpose, we will make use of the perturbative nature of the twist angle under the uniform gauge.
Firstly, we expand the projector in terms of $\theta_j/L_j$ up to the first order near $\boldsymbol{\theta} = 0$:
\begin{eqnarray}
\label{eqn:Projector_expansion_theta_xy}
{\tilde{P}_{\boldsymbol{\theta }}} &=& {\hat{P}} +\frac{{{\theta _x}}}{{{L_x}}}{\left. {\frac{{\partial {{\tilde P}_{\boldsymbol{\theta }}}}}{{\partial \left( {{\theta _x}/{L_x}} \right)}}} \right|_{{\boldsymbol{\theta }} = 0}} + \frac{{{\theta _y}}}{{{L_y}}}{\left. {\frac{{\partial {{\tilde P}_{\boldsymbol{\theta }}}}}{{\partial \left( {{\theta _y}/{L_y}} \right)}}} \right|_{{\boldsymbol{\theta }} = 0}} \nonumber \\
 && + O\left( {\frac{1}{{{L_x}^2}} + \frac{1}{{{L_y}^2}}} \right),
\end{eqnarray}
where we use the fact that $\hat{P}= \tilde{P}_{\boldsymbol{\theta}=0}$ since $\boldsymbol{\theta}=0$ corresponds to general systems under PBC.
Substituiting Eq.~\eqref{eqn:Projector_expansion_theta_xy} into Eq.~\eqref{eqn:ChernNum_PBC_Proj_expression} and keeping up to the leading order lead to:
\begin{equation}
\label{eqn:Berry_curvature_flat}
{{\tilde P}_{\boldsymbol{\theta }}}\left[ {{\partial _{{\theta _x}}}{{\tilde P}_{\boldsymbol{\theta }}},{\partial _{{\theta _y}}}{{\tilde P}_{\boldsymbol{\theta }}}} \right] \approx  {{\hat P}}\left[ {{{\left. {\left( {{\partial _{{\theta _x}}}{{\tilde P}_{\boldsymbol{\theta }}}} \right)} \right|}_{{\boldsymbol{\theta }} = 0}},{{\left. {\left( {{\partial _{{\theta _y}}}{{\tilde P}_{\boldsymbol{\theta }}}} \right)} \right|}_{{\boldsymbol{\theta }} = 0}}} \right] 
\end{equation}
which is independent on the twist angle.
This fact implies that the trace of the Berry curvature defined through TBC should be also independent on the twist angle $\mathrm{Tr}[\tilde{\mathcal F}\left( {\boldsymbol{\theta}} \right)] \approx \mathrm{constant}$ in the thermodynamic limit $L_x, L_y \to \infty$.
We shall examine this point in Sec.~\ref{sec:Numerical_Verification}.

Since the trace of the Berry curvature is a constant for sufficiently large systems, the Chern number [Eq.~\eqref{eqn:ChernNum_PBC_Proj_expression}] can be approximated without implementing integrations \cite{PhysRevLett.122.146601}:
\begin{equation}
 \label{ChernNum_noncommutative_approximated_temp}
{\tilde{C}_{{\mathrm{TBC}}}} \approx - 2\pi i{\mathrm{Tr}}\left( {{{\hat P}}\left[ {{{\left. {\left( {{\partial _{{\theta _x}}}{{\tilde P}_{\boldsymbol{\theta }}}} \right)} \right|}_{{\boldsymbol{\theta }} = 0}},{{\left. {\left( {{\partial _{{\theta _y}}}{{\tilde P}_{\boldsymbol{\theta }}}} \right)} \right|}_{{\boldsymbol{\theta }} = 0}}} \right]} \right).
\end{equation}
Next, we let $\theta_x=2\pi, \theta_y=0$, which maintains the perturbative condition since $2\pi/L_x  \ll 1$.
According to Eq.~\eqref{eqn:P_theta_tilde_relation}, we find the following useful relation
\begin{eqnarray}
\label{eqn:Projector_U2pi_relation}
{{\tilde P}_{\left( {2\pi ,0} \right)}} &=& {{\hat P}} + \frac{{2\pi }}{{{L_x}}}{\left. {\frac{{\partial {{\tilde P}_{\boldsymbol{\theta }}}}}{{\partial \left( {{\theta _x}/{L_x}} \right)}}} \right|_{{\boldsymbol{\theta }} = 0}} + O\left( {\frac{1}{{{L_x}^2}}} \right) \\ \nonumber
 &=& \hat U_{2\pi }^x{{\hat P}}{\left( {\hat U_{2\pi }^x} \right)^{ - 1}},
\end{eqnarray}
where we introduce the twist operator along single direction $\hat U_{{\theta _j}}^j \equiv \exp \left( {i{\theta _j}{{\hat r}_j}/{L_j}} \right),\;j = x,y$ for simplicity.
A similar result can be obtained by letting $\theta_x=0, \theta_y=2\pi$.
Up to the first order of $1/L_j,\;j=x,y$, Eq.~\eqref{eqn:Projector_U2pi_relation} provides a very convenient approach to approximate the partial derivative w.r.t. the twist angle:
\begin{eqnarray}
\label{eqn:approximation_projector_P2pi_P0}
\frac{{2\pi }}{{{L_j}}}{\left. {\frac{{\partial {{\tilde P}_{\boldsymbol{\theta }}}}}{{\partial \left( {{\theta _j}/{L_j}} \right)}}} \right|_{{\boldsymbol{\theta }} = 0}} &=& 2\pi {\left. {\frac{{\partial {{\tilde P}_{\boldsymbol{\theta }}}}}{{\partial {\theta _j}}}} \right|_{{\boldsymbol{\theta }} = 0}} \nonumber \\
&\approx& {{\tilde P}_{2\pi \boldsymbol{e}_j}} - {{\tilde P}}  \nonumber \\
&=& \hat U_{2\pi }^j{{\hat P}}{\left( {\hat U_{2\pi }^j} \right)^{ - 1}} - {{\hat P}}.
\end{eqnarray}
Furthermore, by using the Baker-Campbell-Hausdorff formula, we have
\begin{equation}
{e^{i\frac{{2\pi }}{{{L_j}}}{{\hat r}_j}}}{{\hat P}}{e^{ - i\frac{{2\pi }}{{{L_j}}}{{\hat r}_j}}} = {{\hat P}} + i\frac{{2\pi }}{{{L_j}}}\left[ {{{\hat r}_j},{{\hat P}}} \right] + O\left( {\frac{1}{{{L_j}^2}}} \right),
\end{equation}
and therefore the partial derivative of the projector can be further written as
\begin{equation}
\label{eqn:Projector_partial_derivative_approximation_commutator}
{\left. {\frac{{\partial {{\tilde P}_{\boldsymbol{\theta }}}}}{{\partial {\theta _j}}}} \right|_{{\boldsymbol{\theta }} = 0}} \approx \frac{i}{{{L_j}}}\left[ {{{\hat r}_j},{{\tilde P}}} \right], \quad j=x,y.
\end{equation}
By combining Eqs.~\eqref{eqn:ChernNum_PBC_Proj_expression}, \eqref{eqn:Berry_curvature_flat}, \eqref{eqn:Projector_partial_derivative_approximation_commutator}, we consequently obtain the following real-space form of the Chern number
\begin{eqnarray}
\label{eqn:ChernNum_realspace_noncommutative}
\tilde{C}_{\mathrm{TBC}} &\approx& \frac{{2\pi i}}{{{L_x}{L_y}}}{\mathrm{Tr}}\left( {{{\hat P}}\left[{\left[ {{{\hat r}_x},{{\hat P}}} \right], \left[ {{{\hat r}_y},{{\hat P}}} \right]} \right]} \right) \nonumber \\
 & =& \frac{{ 2\pi i}}{{{L_x}{L_y}}} \sum\limits_{{\boldsymbol{r}},\alpha } {\langle {\boldsymbol{r}},\alpha |\hat P\left[ {\left[ {{{\hat r}_x},\hat P} \right], \left[ {{{\hat r}_y},\hat P} \right]} \right]|{\boldsymbol{r}},\alpha \rangle } .   \nonumber \\
\end{eqnarray}
It can be found that this formula is quite similar to the non-commutative Chern number [Eq.~\eqref{eqn:ChernNum_Noncomm_original}] proposed before (up to a minus sign, depending on the convention of the coordinate).
A minor difference is that Eq.~\eqref{eqn:ChernNum_Noncomm_original} performs trace only within a single cell, while Eq.~\eqref{eqn:ChernNum_realspace_noncommutative} traces for all cells and then averages them.
For clean systems with translation symmetry, one can easily deduce that these two approaches are well consistent.
As for disordered system, they are also consistent if we average over many random realizations.
If the system is inhomogeneous, i.e. a harmonic trap is imposed, it is more reasonable to consider an average over all sites.
We also point out that the approximation in Eq.~\eqref{eqn:approximation_projector_P2pi_P0} is relatively rough in finite systems.
One can use a higher-order finite difference method to gain a more accurate result, see details in \ref{appendix:approximation_finite_difference}.

\subsection{Real-space Chern number via Bott index method}
In this section, we derive the Bott-index form of the real-space Chern number through the TBC.
The Bott index form of the Chern number is proposed to efficiently calculate the intrinsic topological nature in real space \cite{exel1991invariants, Loring_2010, hastings2010almost,HASTINGS20111699, loring2015k, loring2019guide}, and has been applied widely in various systems \cite{PhysRevX.6.011016, PhysRevLett.118.236402,PhysRevLett.121.126401,PhysRevB.98.125130, PhysRevLett.125.217202, PhysRevLett.114.056801}.
It is pointed out that the Bott index is equivalent to the quantum Hall conductance \cite{toniolo2022bott}.
The Bott index is considered as a winding number index of matrices that almost represents the disk or annulus \cite{hastings2010almost, toniolo2022bott}.
For Chern insulators, the Bott index reads as
\begin{equation}
\label{eqn:Bott_Index_original}
{\mathrm{Bott}}\left( {{{\cal U}_x},{{\cal U}_y}} \right) = \frac{1}{{2\pi i}} {{\mathrm{tr}}\left[ {\log \left( {{{\cal U}_x}{{\cal U}_y}{{\cal U}_x}^\dag {{\cal U}_y}^\dag } \right)} \right]},
\end{equation}
where ${\cal U}_{j} ={ \boldsymbol{\Psi}}^\dagger \hat U_{2\pi}^j { \boldsymbol{\Psi}}, j=x,y$.
Here, the notation ``$\mathrm{tr}()$" means the trace operation of the matrix, which implies the trace in the subspace spanned by targeted states.
Next, we show that the Bott index Chern number can be derived from the Chern number defined through TBC [Eq.~\eqref{eqn:ChernNum_TPG}].
The idea is also to use the perturbative nature of the twist angle.
However, to derive a Bott index form, we shall take another route.
Firstly, it can be proved that the Chern number can be written as a winding of Berry phase (line integral) \cite{PhysRevLett.51.51}
\begin{equation}
\label{eqn:Chern_number_Berry_phase_winding}
{\tilde{C}_{{\mathrm{TBC}}}}  = \frac{1}{{2\pi }}\int_0^{2\pi } {{\mathrm{d}}{\theta _x}\;\frac{\partial }{{\partial {\theta _x}}}{\phi^{y} _{{\mathrm{Berry}}}}\left( {{\theta _x}} \right)} 
\end{equation}
in which 
\begin{equation}
{\phi^{y} _{{\mathrm{Berry}}}}\left( {{\theta _x}} \right) = \int_0^{2\pi } {{\mathrm{d}}{\theta _y}\;\mathrm{tr}\left[ {\tilde{\cal A}_y}\left( {{\boldsymbol{\theta}}} \right) \right]} 
\end{equation}
is the Berry phase along the $y$ direction.
Using the perturbative nature of the twist angle, one can obtain the Berry phase in real space via the following expression \cite{PhysRevB.107.125161}
\begin{eqnarray}
{\phi ^y_{{\mathrm{Berry}}}}\left( {{\theta _x}} \right) &=& -i {\mathrm{tr}}\left[ {{\mathrm{exp}}\left( {\int_0^{2\pi } {{\tilde{\cal A}_x}\left( {\boldsymbol{\theta }} \right){\mathrm{d}}{\theta _y}} } \right)} \right]  \nonumber \\
& \approx&  -i\mathrm{tr} \left\{{ \log  \left[ \tilde{\boldsymbol{\Psi }}_{\left( {{\theta _x}, 0} \right)}^\dag {\tilde{\boldsymbol{\Psi }}_{\left( {{\theta _x}, 2\pi} \right)}} \right] }\right\}  \nonumber \\
&=&  -i \mathrm{tr} \left\{{ \log \left[ {{\boldsymbol{\Psi }}_{\left( {{\theta _x}, 0} \right)}^\dag \hat U_{2\pi }^y{{\boldsymbol{\Psi }}_{\left( {{\theta _x}, 0} \right)}}} \right]  }\right\} ,
\end{eqnarray}
where we have used the fact that ${\tilde{\boldsymbol{\Psi }}_{\left( {{\theta _x}, 2\pi} \right)}} = \hat U_{2\pi }^y{\tilde{\boldsymbol{\Psi }}_{\left( {{\theta _x}, 0} \right)}}$.
The above formula can be equivalently formulated via the projected position operator $ \hat{\mathcal{P}}_y(\theta_x) \equiv  {{{\tilde P}_{ (\theta_x, 0)}}{{\hat U}_{2\pi }}^y{{\tilde P}_{ (\theta_x, 0) }}}$ such that
\begin{equation}
\label{eqn:Berry_phase_projected_position_operator}
{\phi _{{\mathrm{Berry}}}}\left( {{\theta _x}} \right) \approx  -i {\mathrm{Tr}}\left[ {\log \left(  \hat{\mathcal{P}}_y(\theta_x) \right)} \right].
\end{equation}
By differentiating Eq.~\eqref{eqn:Berry_phase_projected_position_operator} w.r.t. the twist angle $\theta_x$, we find (see detailed derivation in \ref{appendix:diff_relation})
\begin{equation}
\label{eqn:Berry_phase_diff_relation}
\frac{\partial }{{\partial {\theta _x}}}{\phi _{{\mathrm{Berry}}}}\left( {{\theta _x}} \right) =  -i {\mathrm{Tr}}\left[ {{{\hat {\cal P}}_y}{{({\theta _x})}^\dag }\frac{\partial }{{\partial {\theta _x}}}{{\hat {\cal P}}_y}({\theta _x})} \right].
\end{equation}
Together with Eq.~\eqref{eqn:Chern_number_Berry_phase_winding}, one may find the Chern number is the winding number of projected position operator.
Next, we further treat the twist angle $\theta_x/L_x$ as a perturbation term, and expand the projected position operator $ \hat{\mathcal{P}}_y(\theta_x) $ up to the first order
\begin{equation}
\label{eqn:Projected_position_operator_y_perturbation_expansion}
{\hat {\cal P}_y}\left( {{\theta _x}} \right) = {\hat {\cal P}_y}\left( 0 \right) + {\theta _x}{\left[ {\frac{\partial }{{\partial {\theta _x}}}{{\hat {\cal P}}_y}\left( {{\theta _x}} \right)} \right]_{{\theta _x} = 0}} + O\left( {\frac{1}{{{L_x}^2}}} \right).
\end{equation}
This fact means, up to the first order of $1/L_x$, the Berry phase (polarization) along $y$ direction is a linear function of $\theta_x$, which is consistent with previous studies \cite{PhysRevB.103.075102,PhysRevLett.126.050501, PhysRevLett.126.016402}.
By substituting the expansion in Eq.~\eqref{eqn:Projected_position_operator_y_perturbation_expansion} into Eq.~\eqref{eqn:Berry_phase_projected_position_operator} and integrating the twist angle $\theta_x$, we find the Chern number can be approximated to
\begin{equation}
{\tilde{C}_{{\mathrm{TBC}}}} = -i {\mathrm{Tr}}\left\{ {{\hat{\cal P}_y}{{\left( 0 \right)}^{ \dag}} {{\left[ {\frac{\partial }{{\partial {\theta _x}}}{\hat{\cal P}_y}\left( {{\theta _x}} \right)} \right]}_{{\theta _x} = 0}}} \right\} + O\left( {\frac{1}{{{L_x}^2}}} \right).
\end{equation}
Then, we can make use of the approximation
\begin{equation}
{\hat{\cal P}_y}\left( {{2\pi}} \right) - {\hat{\cal P}_y}\left( {{0}} \right) \approx 2\pi {\left[ {\frac{\partial }{{\partial {\theta _x}}}{\hat{\cal P}_y}\left( {{\theta _x}} \right)} \right]_{{\theta _x} = 0}} ,
\end{equation}
and then approximately we have
\begin{eqnarray}
\label{eqn:ChernNum_TBC_U0U2pi_formula}
{\tilde{C}_{{\mathrm{TBC}}}}  \approx \frac{1}{{2\pi i }}{\mathrm{Tr}}\left[ {{\hat{\cal P}_y}{{\left( 0 \right)}^{ \dag}}{\hat{\cal P}_y}\left( {2\pi } \right) - {\hat{\cal P}_y}{{\left( 0 \right)}^{ \dag}}{\hat{\cal P}_y}\left( {0 } \right) } \right] .
\end{eqnarray}
To obtain the Bott index form, we convert the trace operation ($\mathrm{Tr}$) to the matrix trace ($\mathrm{tr}$) in the subspace spanned by the targeted states 
\begin{eqnarray}
\label{eqn:trace_operation_to_trace_matrix}
&&{\mathrm{Tr}}\left[ {{{\hat {\cal P}}_y}{{\left( 0 \right)}^\dag }{{\hat {\cal P}}_y}\left( {2\pi } \right) - {{\hat {\cal P}}_y}{{\left( 0 \right)}^\dag }{{\hat {\cal P}}_y}\left( 0 \right)} \right]  \nonumber \\
&& \qquad = {\mathrm{tr}}\left( {{{\cal U}_y}^\dag {{\cal U}_x} {{\cal U}_y}{{\cal U}^\dag_x} - {{\cal U}_y}^\dag {{\cal U}_y}} \right).
\end{eqnarray}
In Eq.~\eqref{eqn:trace_operation_to_trace_matrix}, we have used the fact that ${\hat {\cal P}_y}\left( {2\pi } \right) = \hat U_{2\pi }^x{\hat {\cal P}_y}\left( 0 \right){\left( {\hat U_{2\pi }^x} \right)^{ - 1}}$ and $[\hat{U}_{2\pi}^x, \hat{U}_{2\pi}^y] = 0$.
For gapped targeted states, matrices $\mathcal{U}_{x,y}$ are quasi-unitary ${{\cal U}_y}^\dag {{\cal U}_y} \approx I_{\cal N}$ (see proof in \ref{appendix:U_xy_unitary_proof}), in which $I_{\cal N}$ is the $\cal N \times \cal N$ identity matrix in the subspace spanned by the targeted states.
In practical calculations with finite length, one can perform the singular value decomposition (SVD) for these two matrices: ${{\cal U}_{x,y}} = {\boldsymbol{U}}\Sigma {{\boldsymbol{V}}^{ - 1}}$, in which $ {\boldsymbol{U}}, {\boldsymbol{V}} \in \mathrm{U}(\cal{N})$ are unitary matrices, and $\Sigma$ is a diagonal matrix.
The singular values are non-zero and close to identity as long as the targeted states are gapped.
Then, one can enforce all singularity to be identity $\Sigma = I_{\cal N}$, and then the unitarity of the matrix can be guaranteed: ${{\cal U}_{x,y}} = {\boldsymbol{U}} {{\boldsymbol{V}}^{ - 1}} \in \mathrm{U}(\cal{N})$.
Finally, given that $ {{{\cal U}_y}^\dag {{\cal U}_x} {{\cal U}_y}{{\cal U}^\dag_x}}$ is close to the identity matrix, we can utilize the matrix logarithm and approximate Eq.~\eqref{eqn:ChernNum_TBC_U0U2pi_formula} as:
\begin{equation}
\label{eqn:ChernNum_realspace_BottIndex}
{\tilde{C}_{{\mathrm{TBC}}}}  = \frac{1}{{2\pi i}} {{\mathrm{tr}}\left[ {\log \left( {{{\cal U}_y}^\dag {{\cal U}_x} {{\cal U}_y}{{\cal U}_x}^\dag} \right)} \right]} ,
\end{equation}
which is consistent with Eq.~\eqref{eqn:Bott_Index_original}.
Note that the order of these matrices can be circularly permuted in \ref{eqn:trace_operation_to_trace_matrix} due to the trace operation.
The derivation is quite similar to the real-space representation of the winding number in Ref.~\cite{PhysRevB.103.224208}.
Owing to the unitarity of the matrices $\mathcal{U}_{x,y}$, one can conclude that $\det \left( {{{\cal U}_y}^{ \dag}{{\cal U}_x}{{\cal U}_y}{{\cal U}^{\dag}_x}} \right) \in \mathbb{R}$, and therefore the Chern number (Bott index) here should be strictly an integer.

\subsection{Real-space formulae of the spin Chern number in quantum spin Hall insulator}
\label{subsec:QSH}
In this subsection, we focus on the QSH insulator \citep{PhysRevLett.95.146802, bernevig2006quantum} and derive the corresponding real-space topological invariant. 
Unlike the Chern insulator, the QSH insulator obeys the time-reversal symmetry, resulting in a vanishing total Chern number for the gapped band.
Generally speaking, the QSH insulator cannot classified using the conventional Chern number approach.
However, in Ref.~\citep{PhysRevLett.95.226801}, Kane and Mele demonstrate that it can be classified by a  $\mathbb{Z}_2$ quantity instead.
For a spin-1/2 system with well conserved $s_z$, one can faithfully define a spin Chern number.
In this case, the QSH insulator can be simply seen as a two copies of Chern insulators that preserving the time-reversal symmetry.
The spin Chern number can be easily computed in the two different spin sectors in momentum space.
When the conservation of $s_z$ is violated, the two spin sectors become mixed and a well-defined formula for computing the spin Chern number in momentum space is not readily available.
However, the topological property of the QSH insulator should remain as long as the spectral gap is not closed.
To properly capture the topological invariant of QSH insulators, one of the effective methods is to split the projector onto the occupied band into two sectors: $\hat{\mathbb{P}} \left( {\boldsymbol{k}} \right) = {\hat{\mathbb{P}}_ + \left( {\boldsymbol{k}} \right)  } \oplus {\hat{\mathbb{P}}_ - \left( {\boldsymbol{k}} \right) }$ \citep{PhysRevB.80.125327, Prodan_2010}.
Subsequently, the Chern number can be computed in each sector using Eq.~\eqref{eqn:ChernNum_momentum_projector_form} individually.
With this approach, one can derive the real-space formula of the spin Chern number for QSH as the Chern insulator, including the non-commutative method \citep{PhysRevB.80.125327, Prodan_2010}:
\begin{equation}
\label{eqn:QSH_noncommutative_ChernNum}
{C_ \pm } =  - 2\pi i\sum\limits_\alpha  {\langle {{\boldsymbol{r}}_0},\alpha |{{\hat P}_ \pm }\left[ {\left[ {{{\hat r}_x},{{\hat P}_ \pm }} \right],\left[ {{{\hat r}_y},{{\hat P}_ \pm }} \right]} \right]|{{\boldsymbol{r}}_0},\alpha \rangle },
\end{equation}
where $\hat{P}_\pm$ is the spectral projector onto the sector with positive $(+)$ or negative $(-)$ eigenvalues of $\hat{P}\hat{\sigma_z}\hat{P}$.
The Bott index method can be also constructed through this spectral decomposition \citep{PhysRevB.98.125130, PhysRevLett.121.126401}:
 \begin{equation}
 \label{eqn:QSH_BottIndex_ChernNum}
{C_ \pm } = \frac{1}{{2\pi i}}\left\{ {{\rm{tr}}\left[ {\log \left( { V_ \pm ^y V_ \pm ^x V{{_ \pm ^y}^\dag } V{{_ \pm ^x}^\dag }} \right)} \right]} \right\},
 \end{equation}
 with $ V_ \pm ^j = {\hat{P}_ \pm }{e^{i\frac{{2\pi }}{L}{{\hat r}_j}}}{\hat{P}_ \pm } + \left( {I - {\hat{P}_ \pm }} \right),\;j = x,y$.
The spin-Chern number is then identified as $C_{\rm{s}} = (C_+ - C_-)/2$.
On the other hand, it is still possible to define a generalized TBC for the spinful system \citep{PhysRevB.74.045125, PhysRevB.80.125327}:
\begin{equation}
\hat c_{\bold{r}_i}^\dag {t_{\bold{r}_{ij}}}{\hat c_j} \to \left\{ \begin{array}{l}
\hat c_{\bold{r}_i}^\dag {t_{\bold{r}_{ij}}}{e^{i\theta_x{\Gamma _x}}}{{\hat c}_{\bold{r}_j}}\\
\hat c_{\bold{r}_i}^\dag {t_{\bold{r}_{ij}}}{e^{i\theta_y{\Gamma _y}}}{{\hat c}_{\bold{r}_j}}
\end{array} \right., \bold{r}_j\to\bold{r}_i \;{\rm{accros}}\;{\rm{boundary}}
\end{equation}
in which $\hat{c}_{\bold{r}_j}^\dagger=(\hat{c}_{\uparrow, \bold{r}_j}^\dagger, \hat{c}_{\downarrow, \bold{r}_j}^\dagger)$, $t_{ij}$ is $2\times 2$ tunneling matrix, $\theta_{x,y}$ are twist angles, and $\Gamma_{x,y}$ are $2\times 2$ matrices called internal generators.
Such kind of boundary condition is a generalization from the conventional TBC mentioned above.
Then, one can use the TBC formula of Chern number to compute the spin Chern number:
\begin{equation}
{C_{{\mathrm{s}}}} = \frac{1}{{4\pi }}\int\limits^{2\pi }_{0} {\int\limits^{2\pi }_{0} {{\mathrm{Tr}}\left[ {\mathcal{F}\left( {\boldsymbol{\theta }} \right)} \right]} } {\mathrm{d}^2 \boldsymbol{\theta} },
\end{equation}
in which there is an extra $1/2$ factor compared to the general formula of Chern number in Eq.~\eqref{eqn:ChernNum_TBC}.
Similarly, we can introduce the generalized twist operator 
\begin{equation}
\hat U_{\boldsymbol{\theta }}^\Gamma  = \exp \left( {i\frac{{{\theta _x}}}{{{L_x}}}\sum\limits_{\boldsymbol{r}} r_{x}{\hat c_{\boldsymbol{r}}^\dag {\Gamma _x}{{\hat c}_{\boldsymbol{r}}}}  + i\frac{{{\theta _y}}}{{{L_y}}}\sum\limits_{\boldsymbol{r}} r_y{\hat c_{\boldsymbol{r}}^\dag {\Gamma _y}{{\hat c}_{\boldsymbol{r}}}} } \right)
\end{equation}
to transform the boundary term to the bulk
\begin{equation}
\label{eqn:QSH_U_transformation}
\tilde H\left( {\boldsymbol{\theta }} \right) = \hat U_{\boldsymbol{\theta }}^\Gamma \hat H\left( {\boldsymbol{\theta }} \right){\left( {\hat U_{\boldsymbol{\theta }}^\Gamma } \right)^{ - 1}}.
\end{equation}
According to Ref.~\cite{PhysRevB.74.045125}, we can chose $\Gamma_x = \hat{S}$ and $\Gamma_y = 1$ with $\hat{S}$ being the spin operator.
In this case, the system obeys the general TBC along $y$ direction, and the generalized TBC along $x$ direction.
If the system has conserved spin $S$, then Eq.~\eqref{eqn:QSH_U_transformation} will leads to a uniformly distributed gauge field (twist angle).
When $S$ is not conserved, the gauge field is not uniform in $x$ direction.
More precisely, the tunneling matrix will be slightly rotated along $x$ direction, see Appendix.~\ref{appendix:QSH_TBC} for details.
However, as long as the tunneling of particles is finite-range, the gauge field can still be treated as a perturbation.
Following the same route in Sec.~\ref{sec:Chern_number_TBC}, we can replace the position operator by a generalized position operator along $x$ :
\begin{equation}
{{\hat r}_x} \to \hat r_x^\Gamma  \equiv \sum\limits_{\bf{r}} {{r_x}\hat c_{\bf{r}}^\dag {\Gamma _x}{{\hat c}_{\bf{r}}}} ,
\end{equation}
and we leave the $y$ direction unchanged.
Then, we can obtain the non-commutative formula of the spin Chern number:
\begin{equation}
{C_s} = \frac{{\pi i}}{{{L_x}{L_y}}}\sum\limits_{{\bf{r}},\alpha } {\langle {\bf{r}},\alpha |\hat P\left[ {\left[ {{{\hat r}^\Gamma }_x,\hat P} \right],\left[ {{{\hat r}_y},\hat P} \right]} \right]|{\bf{r}},\alpha \rangle } 
\end{equation}
and the Bott index form:
\begin{equation}
{C_{\rm{s}}} = \frac{1}{{4\pi i}}{\rm{tr}}\left[ {\log \left( {{{\cal U}_y}^\dag {\cal U}_x^\Gamma {{\cal U}_y}{\cal U}{{_x^\Gamma }^\dag }} \right)} \right],
\end{equation}
with ${\cal U}_x^\Gamma  = {{\bf{\Psi }}^\dag }\exp \left( {i\frac{{2\pi }}{{{L_x}}}\hat r_x^\Gamma } \right){\bf{\Psi }}$.
It can be noted that these two formulae are slightly different from that in Eq.~\eqref{eqn:QSH_noncommutative_ChernNum} and \eqref{eqn:QSH_U_transformation}.
In fact, the choice of the generalized TBC is not unique.
For example, Refs.~\citep{PhysRevLett.95.136602, PhysRevLett.97.036808} introduce the concept of Chern number matrix based on the TBC.
This will also allow us to establish a real-space formula of the Chern number matrix.
For spin-1/2 QSH insulator, the twist angles, dubbed $\theta^\sigma_{x,y}, \sigma = \uparrow, \downarrow$, are addressed respectively to the two spinful particles when they tunnel through the boundary.
This allows one to define the $2\times 2$ Chern number matrix \citep{PhysRevLett.91.116802, PhysRevLett.97.036808}:
\begin{equation}
\label{eqn:ChernNum_QSH_TBC_Proj_expression}
\mathbb{C}_{\sigma ,\sigma '} = \frac{1}{{2\pi i}}\int\limits_0^{2\pi } {\int\limits_0^{2\pi } {{\rm{Tr}}\left( {{{\hat P}_{\boldsymbol{\theta }}}\left[ {{\partial _{{\theta ^\sigma_x }}}{{\hat P}_{\boldsymbol{\theta }}},{\partial _{{\theta ^{\sigma '}_y}}}{{\hat P}_{\boldsymbol{\theta }}}} \right]} \right)} {{\rm{d}}}{{\theta }_x^{\sigma}}  {{\rm{d}}}{{\theta }_y^{\sigma'}}} ,
\end{equation}
and the TBC spin Chern number can be obtained via \citep{PhysRevLett.97.036808}
\begin{equation}
C_s =\frac{1}{2} \sum\limits_{\sigma ,\sigma '} {{\rm{sgn}} \left( \sigma  \right)\mathbb{C}_{\sigma ,\sigma '}} ,
\end{equation}
where we denote ${\rm{sgn}} \left( \sigma  \right) = +1$ if $\sigma = \uparrow$ and vice versa.
The Chern number matrix method has been proven valid \citep{PhysRevLett.91.116802, PhysRevLett.97.036808, PhysRevB.105.125128}.
Eq.~\eqref{eqn:ChernNum_QSH_TBC_Proj_expression} allows us to follow the strategy in Sec.~\ref{sec:Chern_number_TBC} to derive the real-space formula of the spin Chern number.
Without loss of generality, we shall consider a spin-1/2 system below.
Firstly, we need to introduce position operators in each spin sector: $\hat r_j^\sigma  = \sum\nolimits_{\boldsymbol{r}} {{r_j}\hat n_{\boldsymbol{r}}^\sigma } , j = x,y; \sigma = \uparrow, \downarrow$, and the associated twist operator along $j=x,y$ direction:
\begin{equation}
\hat U_{\theta _j^\sigma }^{j,\sigma } = \exp \left( {i\frac{{\theta _j^\sigma }}{{{L_j}}}\hat r_j^\sigma } \right).
\end{equation}
Under the large gauge transformation, which is similar to Eq.~\eqref{eqn:QSH_U_transformation} with this twist operator, one can also transform the gauge field at the boundary to the bulk.
Although such kind of transformation can not make the gauge field distributed uniformly, particles will only feel a slowly changing gauge field, and the change rate is in the order of $1/L_{x,y}$.
Hence, the perturbative analysis in Sec.~\ref{sec:Perturbative_expansion} can be applied.
With some similar calculations, the real-space version of the spin Chern number matrix reads as
\begin{equation}
\mathbb{C}_{\sigma, \sigma'} = \frac{{\pi i}}{{{L_x}{L_y}}}\sum\limits_{{\bf{r}},\alpha } {\langle {\bf{r}},\alpha |\hat P\left[ {\left[ {{{\hat r}_x^{\sigma} },\hat P} \right],\left[ {{{\hat r}_y^{\sigma'}},\hat P} \right]} \right]|{\bf{r}},\alpha \rangle } ,
\end{equation}
and 
\begin{equation}
\mathbb{C}_{\sigma, \sigma'}  = \frac{1}{{4\pi i}}{\rm{tr}} \left[ \log \left( {{\cal U}^{\sigma'}_y}^\dag {\cal U}_x^{\sigma} {{\cal U}_y}{{\cal U}_x^{\sigma'}}^\dag  \right) \right],
\end{equation}
${\cal U}_j^{\sigma}  = {{\bf{\Psi }}^\dag }\exp \left( {i\frac{{2\pi }}{{{L_j}}}\hat r_j^\sigma } \right){\bf{\Psi }}$, $j=x,y; \sigma = \uparrow, \downarrow$.

\begin{figure}
\centering
  \includegraphics[width = 1 \columnwidth ]{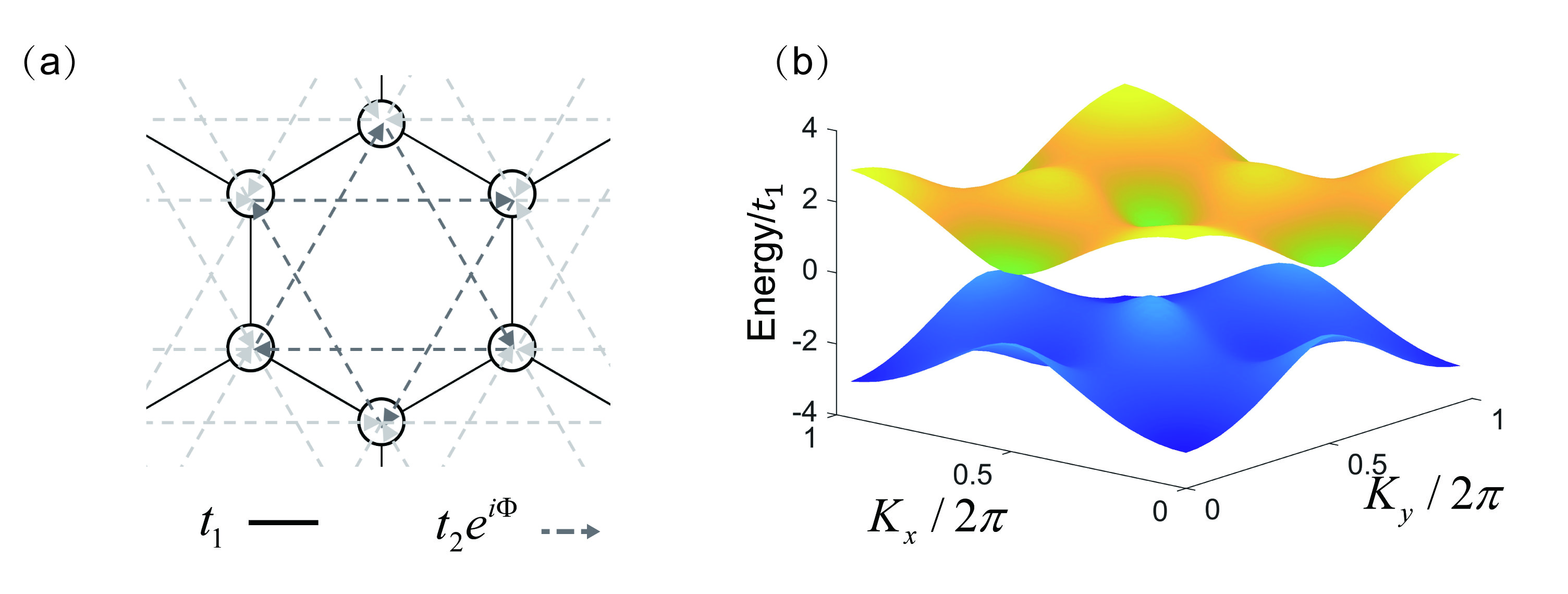}
  \caption{\label{fig:FIG_Haldane_model}
  (a) Schematic illustration of the tunneling relations within a single hexagon in Haldane model and (b) the typical band structure.
  The lattice constant is set to unity for convenience.
    }
\end{figure}

\section{Numerical verifications in the Haldane model and the Kane-Mele model}
\label{sec:Numerical_Verification}
In this section, we present some numerical calculations to verify our results, including the flatness of the Berry curvature defined through TBC and applications of the two real-space formulae of the Chern number.
We shall use the Haldane model to verfiy our result for the Chern insulator, and use the Kane-Mele model for the QSH insulator.

\subsection{The Haldane model}
Here, we consider the celebrated Haldane model \cite{PhysRevLett.61.2015}, which realizes the Chern insulator without requiring net magnetic field.
The Haldane model has been experimentally simulated via ultracold atom platform \cite{jotzu2014experimental, flaschner2018observation}.
The Hamiltonian of Haldane model reads as 
\begin{eqnarray}
\hat H_0 &=&  - \sum\limits_{\langle {\boldsymbol{r}},{\boldsymbol{r}}'\rangle } {{t_1}\hat c_{\boldsymbol{r}}^\dag {{\hat c}_{{\boldsymbol{r}}'}}}  - \sum\limits_{\langle \langle {\boldsymbol{r}},{\boldsymbol{r}}'\rangle \rangle } {{t_2}{e^{i{\Phi _{{\boldsymbol{r}},{\boldsymbol{r}}'}}}}\hat c_{\boldsymbol{r}}^\dag {{\hat c}_{{\boldsymbol{r}}'}}} + \mathrm{H.c.} \nonumber \\
&& + \frac{\Delta_0}{2}\sum\limits_{ {\boldsymbol{r}} } \xi_{\boldsymbol{ r}} {\hat c_{\boldsymbol{r}}^\dag {{\hat c}_{{\boldsymbol{r}}'}}}  
\end{eqnarray}
in which $t_1$ is the nearest-neighbor (NN) tunneling strength, $t_2$ is the next-nearest-neighbor (NNN) tunneling strength, and ${\Phi _{{\boldsymbol{r}},{\boldsymbol{r}}'}}$ is the gauge field addressed to the NNN tunneling, as demonstrated in Fig.~\ref{fig:FIG_Haldane_model}, and $\xi_{\boldsymbol{r}} = \pm 1$ is the energy bias for the two sublattices.
Here $\langle \boldsymbol{r}, \boldsymbol{r}' \rangle$ and $\langle\langle \boldsymbol{r}, \boldsymbol{r}' \rangle\rangle$ respectively represent the NN sites and NNN sites.
%
%

\begin{figure}
\centering
  \includegraphics[width = 1 \columnwidth ]{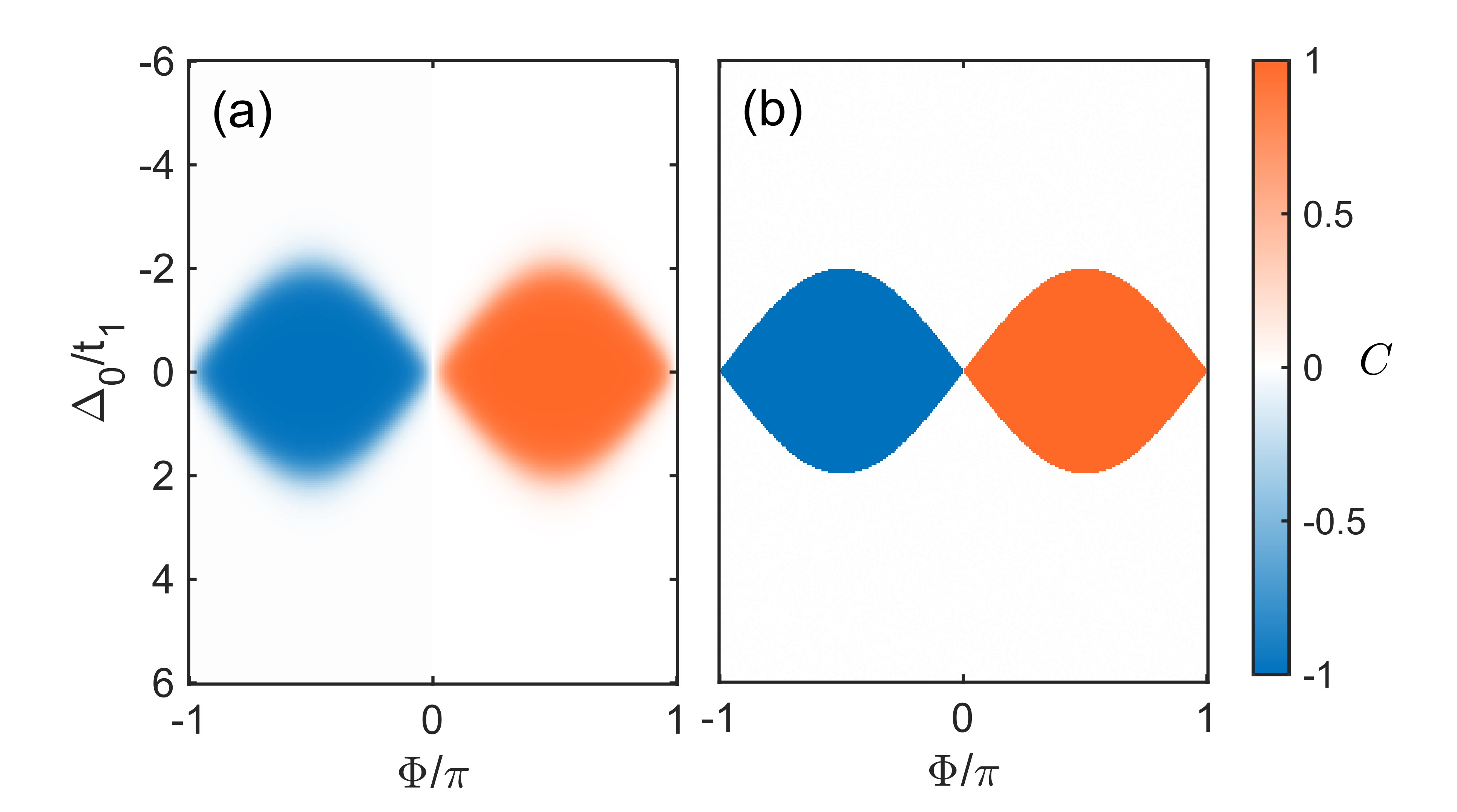}
  \caption{\label{fig:FIG_phase_diagram}
  Phase diagram of the Haldane model.
  (a) and (b) are respectively computed via the non-commutative method [Eq.~\eqref{eqn:ChernNum_realspace_noncommutative}] and the Bott index method [Eq.~\eqref{eqn:ChernNum_realspace_BottIndex}].
  The system's size is chosen as $L_x=L_y = 11$.
  Other parameters: $t_1 = 1, t_2 = 0.2$.
    }
\end{figure}

The phase diagram of the system under PBC is computed through the two real-space formulae, as shown in Fig.~\ref{fig:FIG_phase_diagram} .
The non-commutative method produces relatively smooth results near the phase boundary, while the Bott index method presents a sharp boundary due to the quantized nature.
The phase boundary of the non-commutative method is expected to be sharp in the thermodynamic limit.
Both two real-space formulae can well capture the topological property of the Chern insulator.

\begin{figure}
\centering
  \includegraphics[width = 0.9\columnwidth ]{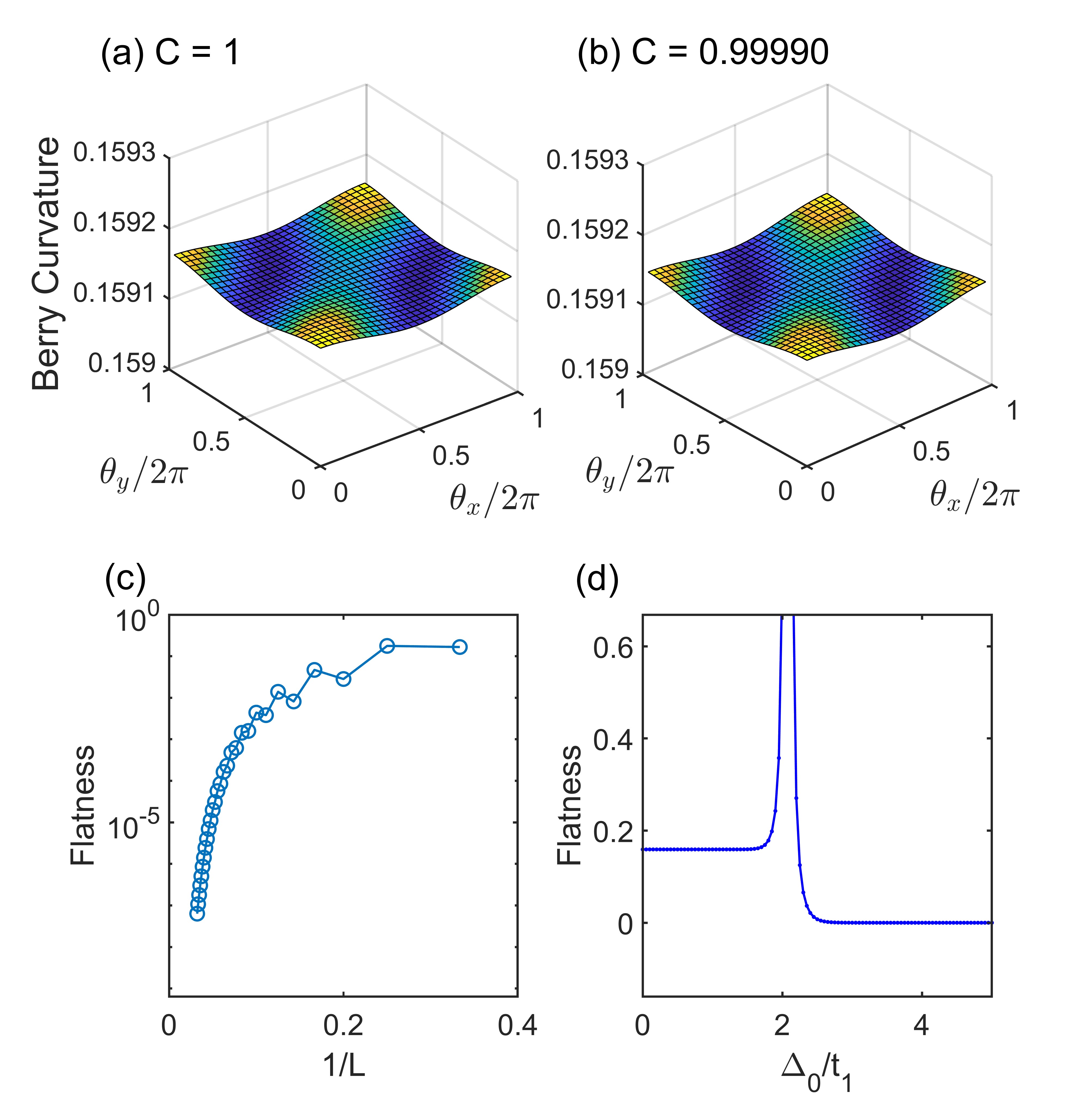}
  \caption{\label{fig:FIG_BerryCurvature_numerical}
  Numerical computation of the Berry curvature defined through the TBC [Eq.~\eqref{eqn:nonAbelian_BerryCurvature_TPG}] with $t_2/t_1 = 0.5$.
  The Chern number is obtained by summing the Berry curvature $C = \frac{1}{{2\pi }}\sum\nolimits_{\boldsymbol{\theta }} {{\mathrm{Tr}}\left[ {\tilde{\cal F}\left( {\boldsymbol{\theta }} \right)} \right]} \delta \theta_x \delta \theta_y $.
  We have used two different numerical methods to compute the Berry curvature:  (a) the link variable method [Eq.~\eqref{eqn:Berry_curvature_link_variable_method}], and (b) the finite difference method [Eq.~\eqref{eqn:Berry_curvature_finite_difference_method}] with $30\times 30$ grid points.
  (c) scaling of the flatness of Berry curvature [Eq.~\eqref{eqn:flatness_ratio}] under different system sizes $L_x=L_y=L$.
  (d) Flatness versus the different sublattice bias energy $\Delta/t_1$ crossing the phase boundary with $t_2/t_1 = 0.2$.
  The topological phase transition occurs at $\Delta_0/t_1 \approx 2$ here.
  The system's size is chosen as $L_x=L_y=11$.
  Other parameters are set to $t_1=1$, $\Phi = \pi/2$. 
  }
\end{figure}

\subsubsection{Flatness of the Berry curvature}
In Sec.~\ref{sec:Perturbative_expansion}, we have shown that the trace of non-Abelian Berry curvature defined through the TBC should tends to be a constant in the thermodynamic limit.
The flat Berry curvature allows us to compute the Chern number without integration, which is important to derive the real-space Chern number.
In this subsection, we shall verify this point numerically.
To compute the Berry curvature numerically through TBC, one should discretize the twist angle and use appropriate numerical method.
Generally, one can adopt the link variable method \cite{JPSJ.74.1674}
\begin{eqnarray}
\label{eqn:Berry_curvature_link_variable_method}
&& {\mathrm{Tr}}\left[ {\tilde{\cal F}\left( {\boldsymbol{\theta }} \right)} \right] \delta \theta_x \delta \theta_y \nonumber \\
 && \qquad =  \ln \det \left[ {{\tilde{\boldsymbol{\Psi }}^\dag }_{({\theta _x},{\theta _y})}{\tilde{\boldsymbol{\Psi }}_{({\theta _x} + \delta {\theta _x},{\theta _y})}}} \right]   \nonumber \\
 && \qquad + \ln \det \left[ {{\tilde{\boldsymbol{\Psi }}^\dag }_{({\theta _x} + \delta {\theta _x},{\theta _y})}{\tilde{\boldsymbol{\Psi }}_{({\theta _x} + \delta {\theta _x},{\theta _y} + \delta {\theta _y})}}} \right]  \nonumber \\
 && \qquad + \ln \det \left[ {{\tilde{\boldsymbol{\Psi }}^\dag }_{({\theta _x} + \delta {\theta _x},{\theta _y} + \delta {\theta _y})}{\tilde{\boldsymbol{\Psi }}_{({\theta _x},{\theta _y} + \delta {\theta _y})}}} \right] \nonumber \\
 && \qquad + \ln \det \left[ {{\tilde{\boldsymbol{\Psi }}^\dag }_{({\theta _x},{\theta _y} + \delta {\theta _y})}{\tilde{\boldsymbol{\Psi }}_{({\theta _x},{\theta _y})}}} \right],
\end{eqnarray}
which produces a gauge-invariant result.
Alternatively, based on Eq.~\eqref{eqn:ChernNum_PBC_Proj_expression}, one can use the finite difference method to approximate the partial derivative
\begin{equation}
{\partial _{{\theta _j}}}{{\tilde P}_{\boldsymbol{\theta }}} \approx \frac{{{{\tilde P}_{{\boldsymbol{\theta }} + \delta {{\boldsymbol \theta }_j}}} - {{\tilde P}_{\boldsymbol{\theta }}}}}{{|\delta {{\boldsymbol \theta }_j}|}},\;\;j = x,y,
\end{equation}
and then we have
\begin{equation}
\label{eqn:Berry_curvature_finite_difference_method}
{\mathrm{Tr}}\left[ {\tilde{\mathcal{F}}\left( {\boldsymbol{\theta }} \right)} \right] \approx {\mathrm{Tr}}\left\{ {{{\tilde P}_{\boldsymbol{\theta }}}\left[ {\frac{{{{\tilde P}_{{\boldsymbol{\theta }} + \delta {{\boldsymbol {\theta} }_x}}} - {{\tilde P}_{\boldsymbol{\theta }}}}}{{|\delta {{\boldsymbol{ \theta} }_x}|}},\frac{{{{\tilde P}_{{\boldsymbol{\theta }} + \delta {{\boldsymbol{ \theta} }_y}}} - {{\tilde P}_{\boldsymbol{\theta }}}}}{{|\delta {{\boldsymbol{ \theta} }_y}|}}} \right]} \right\}.
\end{equation}
It should be noted that the projector ${{\tilde P}_{\boldsymbol{\theta }}}$ is gauge-invariant, and therefore Eq.~\eqref{eqn:Berry_curvature_finite_difference_method} is gauge-invariant as well.
We then numerically compute the Berry curvature using these two methods, as shown in Fig.~\ref{fig:FIG_BerryCurvature_numerical} (a-b).
As expected, the Berry curvature defined through the TBC is very flat.
By summing over all discretized terms, one will obtain the Chern number.
It can be seen that the link variable method [Eq.~\eqref{eqn:Berry_curvature_link_variable_method}] strictly gives a integer Chern number, which is accurate, while the finite difference method [Eq.~\eqref{eqn:Berry_curvature_finite_difference_method}] only produces a Chern number close to integer.
The deviation in the finite difference numerical method is brought by finite grid points.
According to the perturbative expansion in Eq.~\eqref{eqn:Berry_curvature_flat}, the Berry curvature (after tracing) is expected to be a constant in thermodynamic limit.
Next, we examine the flatness of the Berry curvature under different system sizes.
We define the following quantity to reflect the degree of flatness
\begin{equation}
\label{eqn:flatness_ratio}
f = {\max \left\{ {{\mathrm{Tr}}\left[ {\tilde{\cal F}\left( {\boldsymbol{\theta }} \right)} \right]} \right\} - \min \left\{ {{\mathrm{Tr}}\left[ {\tilde{\cal F}\left( {\boldsymbol{\theta }} \right)} \right]} \right\}}.
\end{equation}
The numerical results are shown in Fig.~\ref{fig:FIG_BerryCurvature_numerical} (c).
It can be observed that the flatness of the Berry curvature tends to zero when $L\to \infty$, which is in accordance with our argument.
We then further investigate the flatness of the TBC Berry curvature across topological phase boundary.
As illustrated in Fig.~\ref{fig:FIG_BerryCurvature_numerical} (d), the flatness remains nearly unchanged for various parameter values except at the phase point.
Near the phase boundary, the TBC berry curvature strongly fluctuates.
This observation aligns with our findings, where we demonstrate that the flatness of the Berry curvature solely relies on the system's size.
Near the phase boundary, the spectral gap is extremely small, and the perturbative expansion w.r.t. twist angles up to the first order loses accuracy here, which explains why the flatness becomes parameter-dependent in this regime.

\begin{figure}[htp]
\centering
  \includegraphics[width = \columnwidth ]{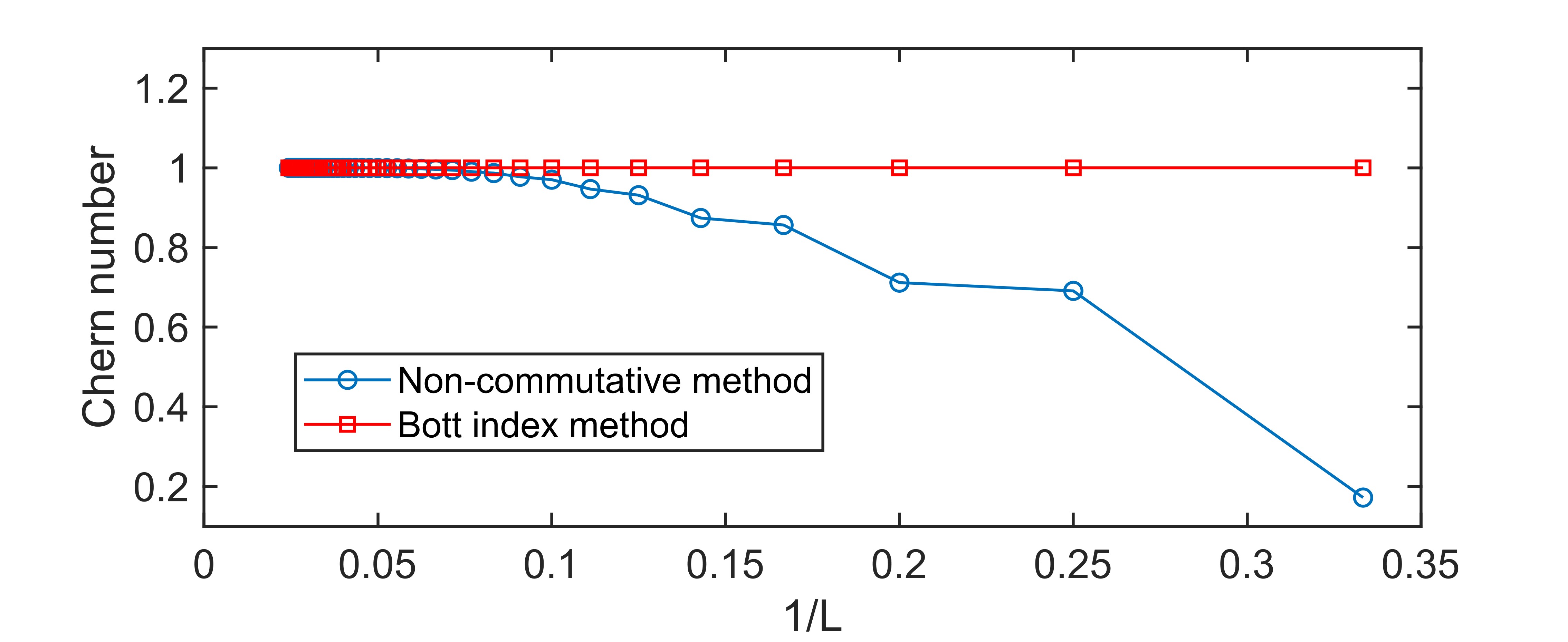}
  \caption{\label{fig:FIG_Scaling_ChernNum}
 Chern number as a function of system size calculated through the two real-space formulae.
  Other parameters are set to: $t_1=1$, $t_2=0.5,\Delta_0 = 0$ and $\Phi = \pi/2$.
    }
\end{figure}

\subsubsection{Real-space Chern number in finite systems}
In Sec.~\ref{sec:Perturbative_expansion}, we have seen that the real-space Chern number is based on the flatness of the Berry curvature defined through TBC.
Nevertheless, the flat Berry curvature is only true in the thermodynamic limit.
In finite systems, the Berry curvature may be dependent on twist angle, and therefore the real-space Chern number may deviate from the correct result.
To see this effect, we compute the Chern number using the real-space formulae [Eq.~\eqref{eqn:ChernNum_realspace_noncommutative} and Eq.~\eqref{eqn:ChernNum_realspace_BottIndex}] with different system sizes.
For finite systems, we follow the idea proposed in Ref.~\cite{PhysRevLett.105.115501, Prodan_2011} to better approximate the finite difference of ${{\tilde P}_{\boldsymbol{\theta }}} $ for the non-commutative Chern number.
Detailed illustration of this higher-order finite difference method is presented in \ref{appendix:approximation_finite_difference}.
Numerical results are shown in Fig.~\ref{fig:FIG_Scaling_ChernNum}.
It can be seen that both methods tends to produce an integer number when $L\to \infty$.
It can be observed that the non-commutative method [Eq.~\eqref{eqn:ChernNum_realspace_noncommutative}] is strongly affected by the system's size.
This is because the non-commutative method relies on the flatness of Berrry curvature.
In Eq.~\eqref{ChernNum_noncommutative_approximated_temp}, we only keep up to the linear terms of $1/L_j$, and regard the Berry curvature as a constant.
Higher-order terms become considerable in finite systems, and the Berry curvature is not necessarily a constant, as already shown in previous subsection.
To gain more accurate results through this method, one may take into account higher terms in the expansion of the projector $\tilde{{P}}_{\boldsymbol{\theta}}$, or increase the system's size.

\begin{figure}[htp]
\centering
  \includegraphics[width = \columnwidth ]{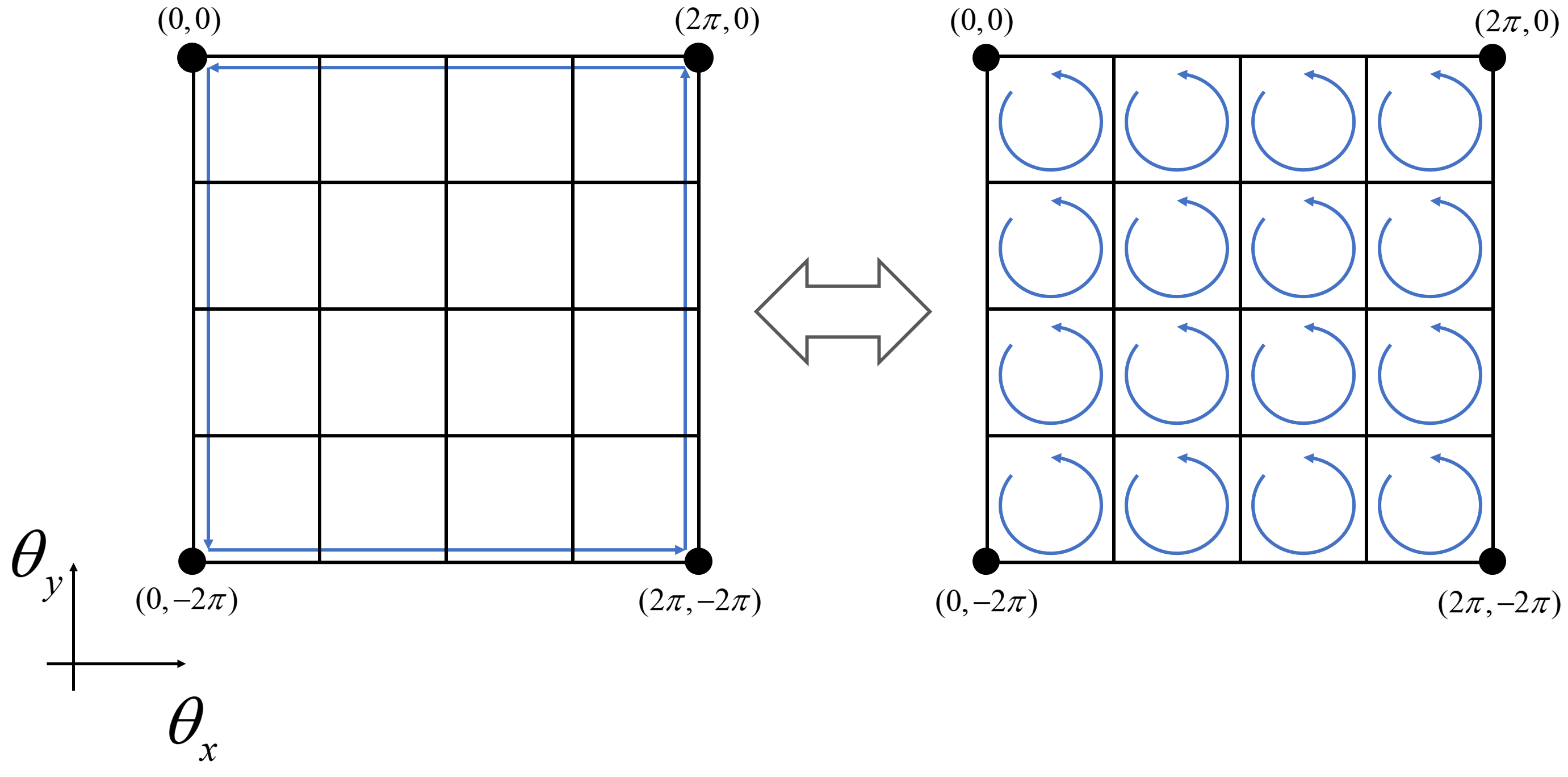}
  \caption{\label{fig:FIG_integral_equivalence}
 The integral of Berry connection (indicated by blue arrows) defined through the TBC along the closed path $\mathcal{D}$ is equivalent to the sum of integral in each small plaquette on the $(\theta_x, \theta_y)$ plane.
    }
\end{figure}

On the other hand, it is surprising to observe that the Bott index method presents a very exact result $C=1$ even for small size.
In other words, the non-flatness of the Berry curvature will not affect the result of the Bott index.
In fact, the Bott index in Eq.~\eqref{eqn:ChernNum_realspace_BottIndex} is approximately related to the following path integral 
\begin{equation}
{{\cal U}_y}^\dag {{\cal U}_x}{{\cal U}_y}{{\cal U}_x}^\dag  \approx \exp \left( {\oint_{\cal D} {\widetilde{\boldsymbol{\mathcal{A}}} \left( {\boldsymbol{\theta }} \right) \cdot {\mathrm{d}}{\boldsymbol{\theta}}} } \right),
\end{equation}
where $\widetilde{\boldsymbol{\mathcal{A}}} \left( {\boldsymbol{\theta }} \right)= ({\tilde{\mathcal{A}}_x}\left( {\boldsymbol{\theta }} \right), {\tilde{\mathcal{A}}_y}\left( {\boldsymbol{\theta }} \right))$, and $\mathcal{D} : (0,0)\to(0, -2\pi) \to (2\pi,-2\pi) \to (2\pi, 0) \to (0, 0)$ is a closed path in $(\theta_x, \theta_y)$ plane.
With Stokes theorem, this integral is equivalent to the sum of the Berry curvature in Eq.~\eqref{eqn:Berry_curvature_link_variable_method}, as depicted in Fig.~\ref{fig:FIG_integral_equivalence}.
This means the Berry curvature is not necessarily flat when applying the Bott index method to calculate the Chern number in real space.
In addition, one may recall that the Stokes theorem fails when Chern number is non-zero in topological band theory.
This is because one can not find a globally smooth gauge for the Bloch state on the Brillouin manifold if the Chern number is non-zero.
Here, we instead consider gapped states $\boldsymbol{\bold{\Psi}}_{\boldsymbol{\theta}}$ parameterized by twist angle and use the non-Abelian form of the Berry connection.
There is no obstruction in applying the Stokes theorem, which explains why the Bott index produces an exactly quantized Chern number.

\subsubsection{Chern number in the presence of disorder}
\begin{figure}
\centering
  \includegraphics[width =  \columnwidth ]{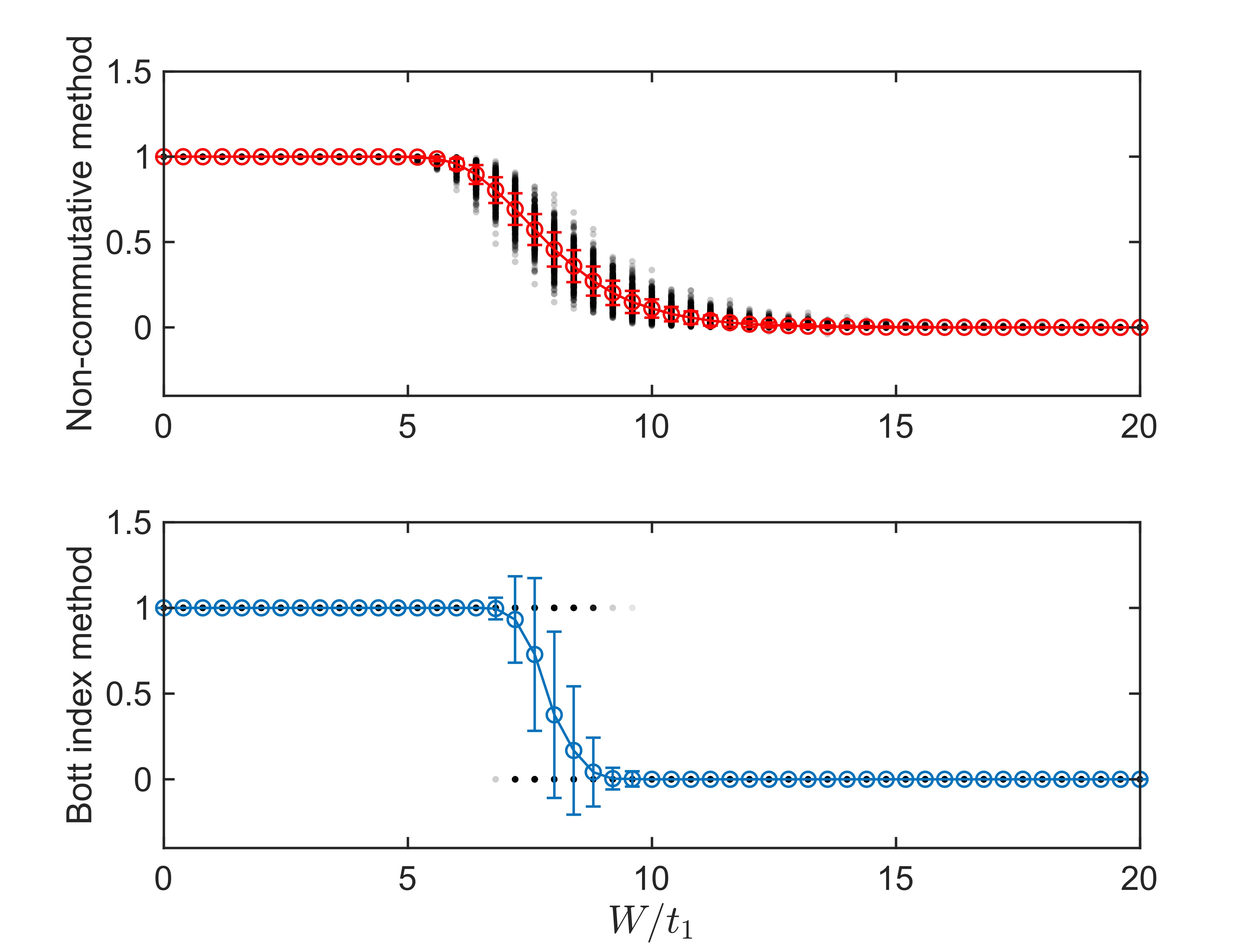}
  \caption{\label{fig:FIG_ChernNumber_disorder}
  Numerical computations of the real-space Chern number as a function of disorder strength $W$ on a $L_x\times L_y = 30\times 30$ system.
  Grey points are from each random realization, while circles are the averaged results.
  We have performed computations over $500$ random realizations.
  The errorbar represents the standard deviations of Chern numbers among these random realizations.
  Other parameters are set to $t_1=1$, $t_2=0.5, \Delta_0 = 0$, $\Phi = \pi/2$. 
  }
\end{figure}

As mentioned in previous sections, the real-space formula allows us to compute the Chern number without requiring the translation symmetry.
To examine this fact, we consider on-site energy disorder on each site's in Haldane model
\begin{equation}
\hat H = {{\hat H}_0} + W\sum\limits_{\boldsymbol{r}} \epsilon _{\boldsymbol{r}}{{\hat n}_{\boldsymbol{r}}}
\end{equation}
where $W$ is the strength of the disorder, and $\epsilon_{\boldsymbol{r}} \in [-0.5, 0.5]$ is a uniformly distributed random number.
We numerically diagonalize the Hamiltonian under PBC and focus on the lowest band.
The parameter is chosen to be $t_1=1$, $t_2=0.5$, $\Phi = \pi/2$, corresponding to the $C=1$ topological phase in the clean system.
Then, we increase the strength of the disorder, and the topological band theory fails in this situation.
By employing the real-space formula (the non-commutative method [Eq.~\eqref{eqn:ChernNum_realspace_noncommutative}] and the Bott index method [Eq.~\eqref{eqn:ChernNum_realspace_BottIndex}], we compute the Chern number as a function of disorder strength $W$, see results in Fig.~\ref{fig:FIG_ChernNumber_disorder}.
The result is averaged over $500$ random realizations.
It can be seen that the Chern number remains $C=1$ for moderate disorder and then there appears a disorder-induced topological transition for stronger disorder.
One can find that results from the two real-space Chern number are well consistent, despite some differences due to the finite-size effect.
As expected, the Bott index method will exactly produce an integer for all realizations, while the non-commutative method will give a non-quantized result near the phase transition point.
These two methods are expected to be equivalent in the thermodynamic limit.

\subsubsection{ Open boundary condition}
\begin{figure}
\centering
  \includegraphics[width =  \columnwidth ]{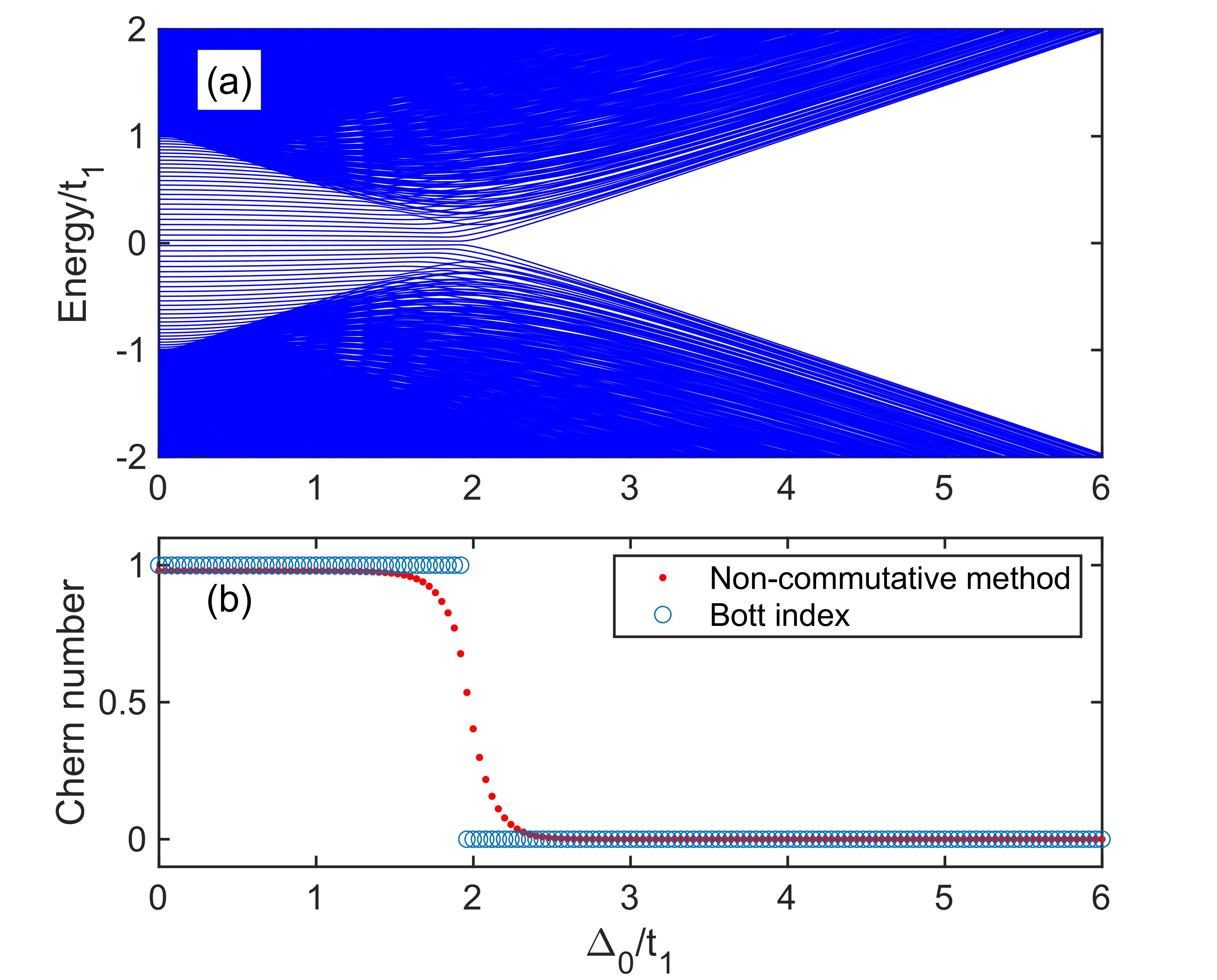}
  \caption{\label{fig:FIG_OBC_ChernNum}
  (a) Enlarged energy spectrum near $E_{\rm{F}} = 0$ as a function of sublattice bias $\Delta_0$ under OBC.
  (b) Chern number as a function of sublattice bias $\Delta_0$ under OBC.
  We set $\delta_L = 3$ to truncate the range of the position operator.
  System's size is chosen to: $L_x = L_y = 30$.
  Other parameters: $t_1 = 1, t_2 = 0.2, \Phi = \pi/2$.
  }
\end{figure}

Previously, we only focus on the PBC or TBC.
In both cases, the system is in a torus geometry and there is no boundary state.
As for the open boundary condition (OBC), there may exist gapless boundary states, bringing problems to numerical computations.
Since these boundary states only localize near the boundary, one can change the definition of the positions operators such that the boundary area is excluded: ${{\hat r}_j} = \sum\nolimits_{{\delta _L} < {{\bf{r}}_j} < {L_j} - {\delta _L}} {{r_j}{{\hat n}_{\bf{r}}}} ,\;j = x,y$.
This trick can circumvent the influence of the boundary state.
In numerical computation, it is sufficient to set $\delta_L$ as a length of few unit cells.
The corresponding system's size shrinks to $L_j'= L_j -2\delta_L, j = x,y$. 
We then compute the Chern number under OBC for different parameters using the reduced position operators, see results in Fig.~\ref{fig:FIG_OBC_ChernNum}.
The Fermi energy is set to $E_{\rm{F}} = 0$ here, and we target those states below it.
It can be clearly seen that the topological phase transition occurs at about $\Delta_0/t_1 \approx 2$, where the in-gap boundary states become absent.
The two real-space formulae can still correctly capture the Chern number.

\subsection{The Kane-Mele model}
\begin{figure}
\centering
  \includegraphics[width =  0.9\columnwidth ]{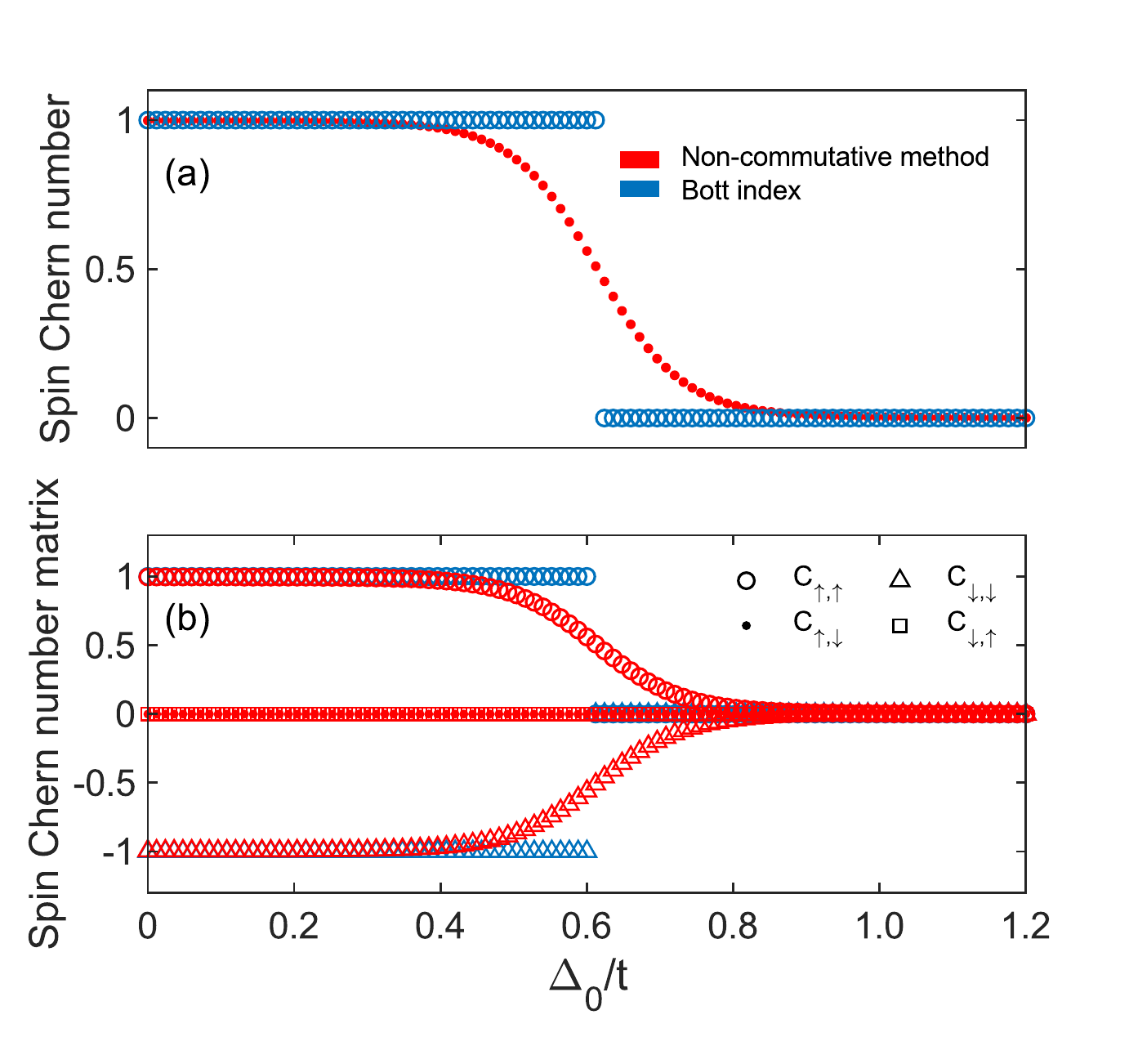}
  \caption{\label{fig:FIG_QSH_Scan_Delta}
  (a) Spin Chern number and (b) elements of spin Chern number matrix calculated through real-space formulae.
  The results of non-commutative method and the Bott index introduced in Sec.~\ref{subsec:QSH} are respectively marked by red and blue colors.
  System size is chosen to: $L_x = L_y = 40$.
  Other parameters: $t = 1$, $\lambda_{\rm{SO}} = 0.06$, $\lambda_{\rm{R}} = 0.015$.
  }
\end{figure}

Now, we briefly discuss the application of the real-space formula of Chern number derived from the generalized TBC for the QSH insulator.
One of the famous model realizing the QSH effect is proposed by Kane and Mele \citep{PhysRevLett.95.226801, PhysRevLett.95.146802}.
The tight-binding Hamiltonian of Kane-Mele model reads as
\begin{eqnarray}
\hat H &= & -t\sum\limits_{\langle ij\rangle } {\hat c_i^\dag {{\hat c}_j}}  + i{\lambda _{{\rm{SO}}}}\sum\limits_{\langle \langle ij\rangle \rangle } {{\nu _{ij}}\hat c_i^\dag \sigma_z {{\hat c}_j}}  \nonumber \\
&& + i{\lambda _{\rm{R}}}\sum\limits_{\langle ij\rangle } {\hat c_i^\dag \left( {{\boldsymbol{\sigma }} \times {{\hat {\boldsymbol{e}}}_{ij}}} \right)_z{{\hat c}_j}}  + \frac{\Delta _0}{2}\sum\limits_i {{\xi _i}\hat c_i^\dag {{\hat c}_i}} ,
\end{eqnarray}
in which $\hat{c}_j^\dag = (\hat{c}^\dag_{\uparrow, j}, \hat{c}^\dag_{\downarrow, j})$, $\lambda_{\rm{SO}}$ is the intrinsic spin-orbit (SO) coupling strength, and $\lambda_{\rm{R}}$ is the Rashba SO coupling strength.
$\nu_{ij} = +1 (-1)$ if the particle tunnels in clockwise (anticlockwise) direction, $\hat{\boldsymbol{e}}_{ij}$ is the normalized vector pointing from $i$ to $j$, and ${\Delta _0}$ is the on-site potential bias.
If $\lambda_{\rm{R}} = 0$, this model simply consists of two copies of Haldane model with opposite NNN tunneling phase $\Phi = \pm \pi/2$, and $s_z$ is conserved.
In this case, one can directly compute the Chern number in each spin sector to attain the spin Chern number.
For $\lambda_{\rm{R}} \ne 0$, the two subspaces are coupled and it becomes a obstacle to compute the spin Chern number in a conventional momentum-space method.
We numerically compute the spin Chern number and the Chern number matrix using the non-commutative method and the Bott index method derived in Sec.~\ref{subsec:QSH}.
As shown in Fig.~\ref{fig:FIG_QSH_Scan_Delta}, the spin Chern number and the Chern number matrix are both capable to produce a correct spin Chern number for the Kane-Mele model.
The non-commutative and the Bott index fomulae derived from the generalized TBC are also confirmed to be valid.

\section{Conclusions and discussions}
\label{sec:Discussion}

In summary, in the thermodynamic limit, we have shown that the non-commutative Chern number and the Bott index form of Chern number can be both derived from the Chern number defined via TBC.
Thus, these two real-space formulae and the TBC formula for the Chern number are equivalent.
Our derivation is based on the perturbative nature of the twist angles under the uniform gauge of TBC.
The key point is to expand the eigenstate and operator to the linear terms of $1/L_{x}$ and $1/L_y$.
The trace of non-Abelian Berry curvature is shown to be a constant in the thermodynamic limit.
Hence, the integration w.r.t. twist angles in the TBC Chern number will be reduced, providing feasibility to deriving the real-space Chern number.
We further derive the real-space formula of spin Chern number throught the generalized TBC.
For different generalized TBCs in QSH insulator, one may obtain different real-space formulae.
Later, the flatness of the trace of Berry curvature defined via TBC is later verified numerically.
The non-flatness effect to the real-space formula of Chern number is discussed.
We also show that the real-space Chern number can be applied to disordered system.
We would like to stress that using the formula of topological invariants defined via TBC to derive the real-space topological invariant may be potentially generalized to other topological systems, such as higher-order topological insulators \citep{PhysRevResearch.2.012009, JPSJ.88.104703, PhysRevLett.128.246602,PhysRevLett.128.127601} and periodically-driven systems in 2D \citep{PhysRevX.3.031005}.
This will benefit the search of real-space formula of topological invariants in unconventional topological insulators.
Throughout this paper, we restrict ourself to the single-particle system.
In fact, the TBC method can be well applied to gapped many-body eigenstates.
Since the real-space formulae of Chern number can be derived from the TBC Chern number, we expect the real-space Chern number can be extended to multi-particle systems.
However, it seems that the Bott index can not work for unique many-body ground state $|\Psi_{\mathrm{GS}}\rangle$ because $\mathcal{U}_{x,y}$ is only a $\mathrm{U}(1)$ number in this case.
Hence, Eq.~\eqref{eqn:ChernNum_realspace_BottIndex} will only produce a trivial result due to the multi-value nature of the exponential function in complex space.
In other words, one can not numerically distinguish trivial and non-trivial topological insulators through the Bott index formula if the ground state is unique.
To circumvent this problem, we recall that the Bott index can be considered as a integral of Berry connection, as mentioned in Fig.~\ref{fig:FIG_integral_equivalence}.
Hence, it is possible to implement the integral for half of the area under TBC with uniform gauge, that is the triangle region $\mathcal{D}' : (0,0)\to(0, -2\pi) \to (2\pi,-2\pi) \to (0, 0)$.
This leads to the following expression
\begin{eqnarray}
{{\tilde C}_{{\rm{TBC}}}} & \approx& \frac{1}{{i\pi }}\log \langle {\tilde{\Psi} _{{\rm{GS}}}}\left( {0,0} \right){\rm{|}}{\tilde{\Psi} _{{\rm{GS}}}}\left( {0, - 2\pi } \right)\rangle  \nonumber \\
&& \qquad \times \langle {\tilde{\Psi} _{{\rm{GS}}}}\left( {0, - 2\pi } \right){\rm{|}}{\tilde{\Psi} _{{\rm{GS}}}}\left( {2\pi, - 2\pi } \right)\rangle  \nonumber \\
&& \qquad \times  \langle {\tilde{\Psi} _{{\rm{GS}}}}\left( {2\pi, - 2\pi } \right){\rm{|}}{\tilde{\Psi} _{{\rm{GS}}}}\left( {0,0} \right)\rangle \nonumber \\
&=& \frac{1}{{i\pi }}\log \langle {\Psi _{{\rm{GS}}}}{\rm{|}}{\left( {\hat U_{2\pi }^y} \right)^{ - 1}}{\rm{|}}{\Psi _{{\rm{GS}}}}\rangle \langle {\Psi _{{\rm{GS}}}}{\rm{|}}\hat U_{2\pi }^x{\rm{|}}{\Psi _{{\rm{GS}}}}\rangle \nonumber \\
&& \qquad \times  \langle {\Psi _{{\rm{GS}}}}{\rm{|}}{\left( {\hat U_{2\pi }^x} \right)^{ - 1}}\hat U_{2\pi }^y{\rm{|}}{\Psi _{{\rm{GS}}}}\rangle ,
\end{eqnarray}
which is quite similar to the real-space marker of the Chern insulator proposed recently in Ref.~\cite{PhysRevB.106.L161106, PhysRevB.107.155106}.
In addition, it should be noted that this quantity also relies on the flatness of the Berry curvature, and it is not necessarily quantized in finite systems.
We only expect it to be quantized in the thermodynamic limit.
As for the non-commutative method [Eq.~\eqref{eqn:ChernNum_realspace_noncommutative}], we can consider a more fundamental formula described in Eq.~\eqref{ChernNum_noncommutative_approximated_temp} for better approximation, which yields
\begin{equation}
{{\tilde C}_{{\mathrm{TBC}}}} \approx  - 2\pi i\langle {\Psi _{{\mathrm{GS}}}}|\left[ {{{\left. {\left( {{\partial _{{\theta _x}}}{{\tilde P}_{\boldsymbol{\theta }}}} \right)} \right|}_{{\boldsymbol{\theta }} = 0}},{{\left. {\left( {{\partial _{{\theta _y}}}{{\tilde P}_{\boldsymbol{\theta }}}} \right)} \right|}_{{\boldsymbol{\theta }} = 0}}} \right]|{\Psi _{{\mathrm{GS}}}}\rangle.
\end{equation}
By using finite difference to approximate the partial derivative, one is also able to obtain an approximated Chern number.

We also note the half-quantized Hall effect has drawn many attentions recently \citep{PhysRevB.105.L201106,PhysRevB.106.035149, PhysRevB.106.045111}.
This is related to gapless surface states, in which the formula of the Chern number [Eq.~\eqref{eqn:ChernNum_TBC}] generally fails.
How to capture the half-quantized nature of Hall conductance through the TBC is challenging.
In such scenario, it is possible to modify the computation method of the Hall conductance.
Then, following the idea in Sec.~\ref{sec:Perturbative_expansion}, one may derive a real-space formula for this system by perturbatively expanding the system up to the first order w.r.t. twist angles.
We believe this could be an intriguing topic in the future.

\acknowledgments
This work is supported by the National Key Research and Development Program of China (Grant No. 2022YFA1404104), and the National Natural Science Foundation of China (Grant No. 12025509, 12247134).

\appendix

\section{Invariance of Chern number under different gauges}
\label{appendix:ChernNumber_boundary_periodic_gauge_equivalence}
We would like to prove that the formulae of TBC Chern number under the boundary gauge and uniform gauge have the same form.
According to Eq.~\eqref{eqn:Ham_PBC_gauge_transform}, there is $ {{\tilde{\boldsymbol{\Psi }}}_{\boldsymbol{\theta }}} = \hat{U}_{\boldsymbol{\theta}}  {{{\boldsymbol{\Psi }}}_{\boldsymbol{\theta }}} $. 
Then, we find the Berry connection is transformed as
\begin{eqnarray}
{{\tilde {\cal A}}_j}({\boldsymbol{\theta }})& =& \tilde {\boldsymbol{\Psi }}_{\boldsymbol{\theta }}^\dag {\partial _{{\theta _j}}}{\tilde {\boldsymbol{\Psi }}_{\boldsymbol{\theta }}} \nonumber \\
 &=& {\boldsymbol{\Psi }}_{\boldsymbol{\theta }}^\dag \hat U_{\boldsymbol{\theta }}^{ - 1}{\partial _{{\theta _j}}}\left( {{{\hat U}_{\boldsymbol{\theta }}}{{\boldsymbol{\Psi }}_{\boldsymbol{\theta }}}} \right) \nonumber \\
 &=& {\boldsymbol{\Psi }}_{\boldsymbol{\theta }}^\dag {\partial _{{\theta _j}}}{{\boldsymbol{\Psi }}_{\boldsymbol{\theta }}} + {\boldsymbol{\Psi }}_{\boldsymbol{\theta }}^\dag \left( {\hat U_{\boldsymbol{\theta }}^{ - 1}{\partial _{{\theta _j}}}{{\hat U}_{\boldsymbol{\theta }}}} \right){{\boldsymbol{\Psi }}_{\boldsymbol{\theta }}} \nonumber \\
 &=& {{\cal A}_j}({\boldsymbol{\theta }})+ {\boldsymbol{\Psi }}_{\boldsymbol{\theta }}^\dag \frac{{{{\hat r}_j}}}{{{L_j}}}{{\boldsymbol{\Psi }}_{\boldsymbol{\theta }}} , \quad j=x, y.
\end{eqnarray}
Hence, it can be found that the Berry curvature satisfies
\begin{equation}
\tilde {\cal F}({\boldsymbol{\theta }}) = {\cal F}({\boldsymbol{\theta }}) + \left[ {{\partial _{{\theta _y}}}{{\bar r}_x}\left( {\boldsymbol{\theta }} \right) - {\partial _{{\theta _x}}}{{\bar r}_y}\left( {\boldsymbol{\theta }} \right)} \right],
\end{equation}
where we denote ${{\bar r}_j}\left( {\boldsymbol{\theta }} \right) \equiv {\boldsymbol{\Psi }}_{\boldsymbol{\theta }}^\dag \frac{{{{\hat r}_j}}}{{{L_j}}}{{\boldsymbol{\Psi }}_{\boldsymbol{\theta }}}$ for simplicity.
It can be seen that generally $\tilde {\cal F}({\boldsymbol{\theta }}) \ne {\cal F}({\boldsymbol{\theta }}) $.
Note that, when the twist angle changes a flux quanta, which is $2\pi$ here, the system returns to the origin.
Meanwhile, it can be immediately checked that ${{\bar r}_j}\left( {\boldsymbol{\theta }} \right) $ is a single-value and gauge-invariant periodic function of twist angle.
Hence, there should be
\begin{equation}
{\rm{Tr}}\left[ {{{\bar r}_j}\left( {{\boldsymbol{\theta }} + 2\pi {{\boldsymbol{e}}_{j'}}} \right)} \right] = {\rm{Tr}}\left[ {{{\bar r}_j}\left( {\boldsymbol{\theta }} \right)} \right], \quad j,j'=x,y,
\end{equation}
where $\bold{e}_{j'}$ represents the unit vector along $j'=x,y$ direction.
After integrating over twist angles, there is
\begin{equation}
\int_0^{2\pi } {\int_0^{2\pi } {\mathrm{d}{\theta _x}\mathrm{d}{\theta _y}\;{\rm{Tr}}\left[ {{\partial _{{\theta _{j'}}}}{{\bar r}_j}\left( {\boldsymbol{\theta }} \right)} \right]} }  = 0, \quad j,j'=x,y.
\end{equation}
Thus, we can conclude that the Chern number defined through TBC is the same under the boundary gauge and the uniform gauge.
In other words, the large gauge transformation in Eq.~\eqref{eqn:Ham_PBC_gauge_transform} will not affect the Chern number.

\section{Approximation for finite difference}
\label{appendix:approximation_finite_difference}
To better approximate the partial derivative of the projector matrix numerically, one can use a higher-order finite difference method for this first-order derivative introduced in Ref.~\citep{PhysRevLett.105.115501, Prodan_2011}.
To see how it works, we write down series expansion of the projector with respect to $\theta/L$:
\begin{equation}
{P_{\theta /L}} = {P_0} + \sum\limits_{k = 1}^\infty  {{{\left( {\frac{\theta }{L}} \right)}^k}\frac{1}{{k!}}} {\left[ {\frac{{{\partial ^k}}}{{\partial {{\left( {\theta /L} \right)}^k}}}{P_{\theta/L} }} \right]_{\theta  = 0}}.
\end{equation}
As a demonstration, we assume that the projector is only parameterized by $\theta$.
The result can be readily extended to the 2D system discussed in the main text.
By construction, there is
\begin{eqnarray}
\frac{ {P_{\theta /L}} - {P_{ - \theta /L}}}{2} &=& \sum\limits_{m = 1}^\infty  {{{\left( {\frac{\theta }{L}} \right)}^{2m - 1}}\frac{1}{{\left( {2m - 1} \right)!}}} \nonumber \\
&&\quad \times {\left[ {\frac{{{\partial ^{2m - 1}}}}{{\partial {{\left( {\theta /L} \right)}^{2m - 1}}}}{P_{\theta/L} }} \right]_{\theta  = 0}}.
\end{eqnarray}
Thus, by choosing a set of parameters $\{ c_n \}$ to eliminate higher-order terms, the first-order derivative can be written as
\begin{equation}
{\frac{\theta }{L}} {\left[ {\frac{\partial }{{\partial \left( {\theta /L} \right)}}{P_{\theta /L}}} \right]_{\theta  = 0}} \approx \sum\limits_{n\in\mathbb{Z^{+}}} {\frac{c_n}{2}\left( {{P_{n\theta /L}} - {P_{ - n\theta /L}}} \right)} 
\end{equation}
The coefficient $\{ c_n \}$ can be solved through a set of linear equations according to the series expansion.
By truncating the expansion to $Q$ terms with $Q<L/2$ (to ensure the convergence of the expansion), it can be written in a matrix form
\begin{eqnarray}
&& \left( {\begin{array}{*{20}{c}}
{{c_1}}&{{c_2}}& \cdots &{{c_Q}}
\end{array}} \right)\left( {\begin{array}{*{20}{c}}
{{P_{\theta /L}} - {P_{ - \theta /L}}}\\
{{P_{2\theta /L}} - {P_{ - 2\theta /L}}}\\
 \vdots \\
{{P_{Q\theta /L}} - {P_{ - Q\theta /L}}}
\end{array}} \right)  \times \frac{1}{2}  \nonumber \\
&& \qquad\qquad\qquad = \frac{\theta }{L}{\left[ {\frac{\partial }{{\partial \left( {\theta /L} \right)}}{P_{\theta /L}}} \right]_{\theta  = 0}}.
\end{eqnarray}
Then, we can convert this problem to solving a set of linear equations, which can be expressed as
\begin{equation}
\left( {\begin{array}{*{20}{c}}
1&2& \cdots &Q\\
{{1^3}}&{{2^3}}& \cdots &{{Q^3}}\\
 \vdots & \vdots & \ddots & \vdots \\
{{1^{2Q - 1}}}&{{2^{2Q - 1}}}& \cdots &{{Q^{2Q - 1}}}
\end{array}} \right)\left( {\begin{array}{*{20}{c}}
{{c_1}}\\
{{c_2}}\\
 \vdots \\
{{c_Q}}
\end{array}} \right) = \left( {\begin{array}{*{20}{c}}
1\\
0\\
 \vdots \\
0
\end{array}} \right)
\end{equation}
Hence, by taking the matrix inverse, $\{ c_n \}$ can be solved numerically through:
\begin{equation}
\left( {\begin{array}{*{20}{c}}
{{c_1}}\\
{{c_2}}\\
 \vdots \\
{{c_Q}}
\end{array}} \right) = {\left( {\begin{array}{*{20}{c}}
1&2& \cdots &Q\\
{{1^3}}&{{2^3}}& \cdots &{{Q^3}}\\
 \vdots & \vdots & \ddots & \vdots \\
{{1^{2Q - 1}}}&{{2^{2Q - 1}}}& \cdots &{{Q^{2Q - 1}}}
\end{array}} \right)^{ - 1}}\left( {\begin{array}{*{20}{c}}
1\\
0\\
 \vdots \\
0
\end{array}} \right).
\end{equation}
Next, by using the fact that
\begin{equation}
{P_{2n\pi /L}} = {\left( {{{\hat U}_{2\pi }}} \right)^n}{P_{0}}{\left( {{{\hat U}_{ - 2\pi }}} \right)^n}
\end{equation}
we can let $\theta = 2\pi$ and then:
\begin{eqnarray}
&&\frac{{2\pi }}{L}{\left[ {\frac{\partial }{{\partial \left( {\theta /L} \right)}}{P_{\theta /L}}} \right]_{\theta  = 0}} \nonumber \\
&& \;\;  \approx \sum\limits_{n=1}^{Q} {\frac{c_n}{2}\left[ {{{\left( {{{\hat U}_{2\pi }}} \right)}^n}{P_0}{{\left( {{{\hat U}_{ - 2\pi }}} \right)}^n} - {{\left( {{{\hat U}_{ - 2\pi }}} \right)}^n}{P_0}{{\left( {{{\hat U}_{2\pi }}} \right)}^n}} \right]} ,   \nonumber \\
\end{eqnarray}
which can be evaluated numerically and produces a good approximation for the partial derivative.

\section{Quasi-unitarity of the matrices $\mathcal{U}_{x,y}$}
\label{appendix:U_xy_unitary_proof}
Below, we show that the matrix $\mathcal{U}_j =  {\boldsymbol{\Psi }}_{0}^\dag  \hat{U}_{2\pi}^{j}  {\boldsymbol{\Psi }}_{0}, j=x,y$ mentioned in the main text is a unitary matrix in the thermodynamic limit $L_j\to \infty$ through the TBC.
Firstly, note that 
\begin{equation}
\mathcal{U}_j  = {\boldsymbol{\Psi }_0^\dag}{\hat{U}^j_{2\pi }}{{\boldsymbol{\Psi }_0} }, \qquad {\mathcal{U}_j ^\dag } = {\boldsymbol{\Psi }_0^{\dag}}({\hat{U}^j_{2\pi }})^{ - 1}{{\boldsymbol{\Psi }_0} }.
\end{equation}
The vector is normalized: ${\boldsymbol{\Psi }_0^{\dag}}{{\boldsymbol{\Psi }_0} }  = {I_{\mathcal N}}$, and $\mathcal N$ is the number of targeted states.
Meanwhile, we have ${\boldsymbol{\Psi }_0}{{\boldsymbol{\Psi }_0^{\dag}} } = {\sum\nolimits_{\mu \in \rm{target}}  {|{{ \psi }_\mu }(0)\rangle \langle {{ \psi }_\mu }(0)|} } = 1$ in the subspace spanned by targeted states without the TBC.
There is
\begin{eqnarray}
{\mathcal{U}_{2\pi}^j}{\left({\mathcal{U}}^j_{2\pi}\right)^\dag } &=& {\boldsymbol{\Psi }_0^\dag}{\hat{U}^j_{2\pi }}{{\boldsymbol{\Psi }_0} }{\boldsymbol{\Psi }_0^{\dag}}(\hat{U}^j_{2\pi })^{ - 1}{{\boldsymbol{\Psi }_0} } \nonumber \\
 &=& {\boldsymbol{\Psi }_0^\dag}{\hat{U}^j_{2\pi }}\left( {\sum\limits_\mu  {|{{ \psi }_\mu(0) }\rangle \langle {{ \psi }_\mu } (0) |} } \right)(\hat{U}^j_{2\pi })^{ - 1}{{\boldsymbol{\Psi }_0} } \nonumber \\
 &=& {\boldsymbol{\Psi }_0^\dag}\left( {\sum\limits_\mu  {|{\tilde{\psi} _\mu }\left( {2\pi \boldsymbol{e}_j } \right)\rangle \langle {\tilde{\psi} _\mu }\left( {2\pi\boldsymbol{e}_j} \right)|} } \right){{\boldsymbol{\Psi }_0} }.\nonumber \\
\end{eqnarray}
Approximately, there is
\begin{eqnarray}
\sum\limits_\mu  {|{\psi _\mu }\left( \boldsymbol{\theta} \right)\rangle \langle {\psi _\mu }\left( {\boldsymbol{\theta} } \right)|}  &=& \sum\limits_\mu  {|{{ \psi }_\mu }\rangle \langle {{ \psi }_\mu }|}  + O\left( {\frac{1}{L}}  \right)  \nonumber \\
&=& 1  + O\left( {\frac{1}{L}}  \right) 
\end{eqnarray}
which implies that $\sum_{\mu \in \rm{target}}  {|{\psi _\mu }\left( {2\pi } \right)\rangle \langle {\psi _\mu }\left( {2\pi } \right)|} $ is close to the identity matrix in this subspace.
Hence, we find ${\mathcal M}{{\mathcal M}^\dag } = {I_{\mathcal N}}$ in the thermodynamic limit.
One can also prove that ${{\mathcal M}^\dag }{\mathcal M} = {I_{\mathcal N}}$ using the same analysis.
In summary, we have shown that the matrix ${\mathcal M}$ is approximately a unitary matrix in the thermodynamic limit.
Similar conclusions can be found in Refs.~\citep{Loring_2010,hastings2010almost}.

\section{Derivations of Eq.~\eqref{eqn:Berry_phase_diff_relation}}
\label{appendix:diff_relation}
Here we show that how to derive the partial derivative of the Berry phase formulated by projected position operator.
The projected position operator reads as
\begin{eqnarray}
&&{{\tilde P}_{\left( {{\theta _x},0} \right)}}{{\hat U}_{2\pi }}^y{{\tilde P}_{\left( {{\theta _x},0} \right)}} \nonumber \\
&& \qquad = \sum\limits_{\mu,\mu'  \in \rm{target}} {{{\left[ {{{\cal U}_y}\left( {{\theta _x}} \right)} \right]}_{\mu ,\mu '}}|{\psi _\mu }\left( {{\theta _x}} \right)\rangle \langle {\psi _{\mu '}}\left( {{\theta _x}} \right)|} . \nonumber \\
\end{eqnarray}
One can diagonalize the matrix ${\cal U}_y (\theta_x)$ to find
\begin{equation}
{{\tilde P}_{\left( {{\theta _x},0} \right)}}{{\hat U}_{2\pi }}^y{{\tilde P}_{\left( {{\theta _x},0} \right)}} = \sum\limits_n {{e^{i{\Phi _n}\left( {{\theta _x}} \right)}}|{\Phi _n}\left( {{\theta _x}} \right)\rangle \langle {\Phi _n}\left( {{\theta _x}} \right)|} ,
\end{equation}
where we assumed ${{{\cal U}_y}\left( {{\theta _x}} \right)}$ is unitary and hence its eigenvalues are $\rm{U}(1)$ numbers.
Then, the logarithm of the projected position operator can be equivalently expressed as
\begin{equation}
\log \left[ {{\tilde P}_{\left( {{\theta _x}, 0} \right)}}{{\hat U}_{2\pi }}^y{{\tilde P}_{\left(  {{\theta _x}, 0} \right)}} \right] = i\sum\limits_n {{\Phi _n}\left( {{\theta _x}} \right)|{\Phi _n}\left( {{\theta _x}} \right)\rangle \langle {\Phi _n}\left( {{\theta _x}} \right)|} .
\end{equation}
Taking the partial derivative yields
\begin{eqnarray}
&& \frac{\partial }{{\partial {\theta _x}}}\log \left[ {{{\tilde P}_{\left(  {{\theta _x}, 0} \right)}}\hat U_{2\pi }^x{{\tilde P}_{\left(  {{\theta _x}, 0} \right)}}} \right] \nonumber  \\
&& \qquad= i\sum\limits_n {\left[ {\frac{\partial }{{\partial {\theta _x}}}{\Phi _n}\left( {{\theta _x}} \right)} \right]|{\Phi _n}\left( {{\theta _x}} \right)\rangle \langle {\Phi _n}\left( {{\theta _x}} \right)|} \nonumber  \\
 && \;\; \qquad+ i\sum\limits_n {{\Phi _n}\left( {{\theta _x}} \right)\frac{\partial }{{\partial {\theta _x}}}\left[ {|{\Phi _n}\left( {{\theta _x}} \right)\rangle \langle {\Phi _n}\left( {{\theta _x}} \right)|} \right]}. 
\end{eqnarray}
The second term will vanish under the trace operation, and then we have
\begin{equation}
{\rm{Tr}}\left\{ {\frac{\partial }{{\partial {\theta _x}}}\log \left[ {{{\tilde P}_{\left( {0,{\theta _x}} \right)}}\hat U_{2\pi }^x{{\tilde P}_{\left( {0,{\theta _x}} \right)}}} \right]} \right\} = i\sum\limits_n {\frac{\partial }{{\partial {\theta _x}}}{\Phi _n}\left( {{\theta _x}} \right)} .
\end{equation}
Note that, in band insulator, ${\Phi _n}\left( {{\theta _x}} \right)$ is proportional to the center of Wannier function.
In the same vein, one can find that
\begin{eqnarray}
&& {\rm{Tr}}\left\{ {\left[ {{{\tilde P}_{\left(  {{\theta _x}, 0} \right)}}{{\left( {\hat U_{2\pi }^y} \right)}^{ - 1}}{{\tilde P}_{\left( {{\theta _x}, 0} \right)}}} \right]\frac{\partial }{{\partial {\theta _x}}}\left[ {{{\tilde P}_{\left( {0,{\theta _x}} \right)}}\hat U_{2\pi }^y{{\tilde P}_{\left(  {{\theta _x}, 0} \right)}}} \right]} \right\}  \nonumber \\
&& \qquad = i\sum\limits_n {\frac{\partial }{{\partial {\theta _x}}}{\Phi _n}\left( {{\theta _x}} \right)}  \nonumber \\
&& \qquad = {\rm{Tr}}\left\{ {\frac{\partial }{{\partial {\theta _x}}}\log \left[ {{{\tilde P}_{\left(  {{\theta _x}, 0} \right)}}\hat U_{2\pi }^y{{\tilde P}_{\left(  {{\theta _x}, 0} \right)}}} \right]} \right\},
\end{eqnarray}
which gives rise to Eq.~\eqref{eqn:Berry_phase_diff_relation}.

\section{Gauge transformation for the generalized TBC}
\label{appendix:QSH_TBC}
Note that the twist operator manifests that
\begin{equation}
\hat U_{\boldsymbol{\theta }}^\Gamma \hat c_i^\dag {\left( {\hat U_{\boldsymbol{\theta }}^\Gamma } \right)^{ - 1}} = \hat c_i^\dag {{\mathop{\rm e}\nolimits} ^{i\left( {\frac{{{r_x}{\theta _x}}}{{{L_x}}}{\Gamma _x} + \frac{{{r_y}{\theta _y}}}{{{L_y}}}{\Gamma _y}} \right)}}.
\end{equation}
Therefore, the tunneling term will be transformed to
\begin{eqnarray}
\hat U_{\boldsymbol{\theta }}^\Gamma \hat c_{{{\boldsymbol{r}}_i}}^\dag {t_{ij}}{{\hat c}_{{{\boldsymbol{r}}_j}}}{\left( {\hat U_{\boldsymbol{\theta }}^\Gamma } \right)^{ - 1}} = \hat c_{{{\boldsymbol{r}}_i}}^\dag {{\tilde t}_{ij}}\left( {\boldsymbol{\theta }} \right){{\hat c}_{{{\boldsymbol{r}}_j}}}
\end{eqnarray}
where the tunneling matrix is rotated
\begin{equation}
{{\tilde t}_{\bold{r}_{ij}}}\left( {\boldsymbol{\theta }} \right) = {{\mathop{\rm e}\nolimits} ^{i\left( {\frac{{{r_{{x_i}}}{\theta _x}}}{{{L_x}}}{\Gamma _x} + \frac{{{r_{{y_i}}}{\theta _y}}}{{{L_y}}}{\Gamma _y}} \right)}}{t_{\bold{r}_{ij}}}{{\mathop{\rm e}\nolimits} ^{ - i\left( {\frac{{{r_{{x_j}}}{\theta _x}}}{{{L_x}}}{\Gamma _x} + \frac{{{r_{{y_j}}}{\theta _y}}}{{{L_y}}}{\Gamma _y}} \right)}}.
\end{equation}
For $\Gamma_x = S$, $\Gamma_y = 1$, the tunneling matrix becomes uniform along $y$ direction and spatially modulated along $x$ direction if $[S, t_{\bold{r}_{ij}}] \ne 0$.
The rotation degree is in order of $1/L_x$ between two adjacent sites provided the tunneling is finite-range.
%

\bibliography{RealSpace_ChernNum_bib}

\end{document}